\documentclass[universe,article,accept,pdftex,moreauthors]{mdpi} 

\usepackage{subfigure}
\usepackage{epstopdf}
\usepackage{color}

\firstpage{1} 
\pubvolume{8}
\issuenum{1}
\articlenumber{182}
\pubyear{2022}
\copyrightyear{2022}
\externaleditor{Academic Editor: Luca Buoninfante and Gaetano Lambiase}
\datereceived{2 February 2022} 
\dateaccepted{9 March 2022} 
\datepublished{14 March 2022} 
\hreflink{https://doi.org/10.3390/universe8030182} 

\newcommand{\ba}{\begin{eqnarray}}
\newcommand{\ea}{\end{eqnarray}}
\newcommand{\be}{\begin{equation}}
\newcommand{\ee}{\end{equation}}

\usepackage{amssymb}
\usepackage{soul}
\usepackage{color}

\def\be{\begin{equation}}
\def\ee{\end{equation}}
\def\bea{\begin{eqnarray}}
\def\eea{\end{eqnarray}}

\Title{Optical Features of AdS Black Holes in the Novel 4D Einstein-Gauss-Bonnet Gravity Coupled to Nonlinear Electrodynamics}
\TitleCitation{Optical Features of AdS Black Holes in the Novel 4D Einstein-Gauss-Bonnet Gravity Coupled to Nonlinear Electrodynamics}

\Author{Khadije Jafarzade $^{1}$, Mahdi Kord Zangeneh $^{2}$ and Francisco S.N. Lobo $^{3,4,}$*}
\AuthorNames{Khadije Jafarzade, Mahdi Kord Zangeneh and Francisco S.N. Lobo}
\AuthorCitation{Jafarzade, K.; Zangeneh, M.K.; Lobo, F.S.N.}

\address{%
\textsuperscript{1} \quad Department of Physics, Faculty of Science, University of Mazandaran, Babolsar 47416-95447, Iran; khadije.jafarzade@gmail.com\\
\textsuperscript{2} \quad Physics Department, Faculty of Science, Shahid Chamran University of Ahvaz, Ahvaz 61357-43135, Iran; mkzangeneh@scu.ac.ir\\
\textsuperscript{3} \quad Instituto de Astrof\'{i}sica e Ci\^{e}ncias do Espa\c{c}o, Faculdade de Ci\^encias da Universidade de Lisboa, \linebreak Edif\'{i}cio C8, Campo Grande, PT1749-016 Lisbon, Portugal\\
\textsuperscript{4} \quad Departamento de F\'{i}sica, Faculdade de Ci\^{e}ncias, Universidade de Lisboa, Edifício C8, Campo Grande, PT1749-016 Lisbon, Portugal}

\corres{Correspondence: fslobo@fc.ul.pt; Tel.: +351-217-500-484; Fax: 351-217-500-977}

\abstract{An alternative theory of gravity that has attracted much attention recently is the novel four-dimensional Einstein-Gauss-Bonnet (4D EGB) gravity. The theory is rescaled by the Gauss-Bonnet (GB) coupling constant $\alpha \rightarrow \alpha/(D - 4)$ in $ D $ dimensions and redefined as four-dimensional gravity in the limit $D \rightarrow 4$. Thus, in this manner, the GB term yields a non-trivial contribution to the gravitational dynamics.
In fact, regularized black hole solutions and applications in the novel 4D EGB gravity have also been extensively explored. In this work, motivated by recent astrophysical observations, we present an in-depth study of the optical features of AdS black holes in the novel 4D EGB gravity coupled to exponential nonlinear electrodynamics (NED), such as the shadow geometrical shape, the energy emission rate, the deflection angle and quasinormal modes. 
Taking into account these dynamic quantities, we investigate the effects on the black hole solution by varying the parameters of the models. More specifically, we show that the variation of the GB and NED parameters, and of the cosmological constant, imprints specific signatures on the optical features of AdS black holes in the novel 4D EGB gravity coupled to nonlinear electrodynamics, thus leading to the possibility of directly testing these black hole models by using astrophysical observations.
}


\begin{document}

\section{Introduction}

A natural generalization of general relativity is Lovelock gravity, which lead to second order differential equations for the metric functions. The~simplest Lovelock theory of gravity is well-known as Gauss-Bonnet (GB) gravity, which contains higher curvature terms in the action and encompasses non-trivial dynamics for higher-dimensional $(D > 4)$ theories of gravity. Nevertheless, in~a four-dimensional description, the~GB term is topological invariant, and~hence does not contribute to the gravitational dynamics. To~generate a nontrivial contribution, one usually couples the GB term to a scalar field. Recently, Glavan and Lin suggested that by
rescaling the GB coupling constant $\alpha \rightarrow \alpha /(D-4)$ and defining the
four-dimensional theory as the limit $D \rightarrow 4$, the~GB term can yield a non-trivial contribution to the gravitational dynamics and the theory can bypass the Lovelock theorem~\cite{EGB0}. The~theory, now dubbed as the novel 4D Einstein-Gauss-Bonnet (EGB) gravity, is free from the Ostrogradsky instability~\cite{EGB01}. 

In fact, the~novel 4D EGB theory has attracted extensive attention recently, ranging from applications of black hole physics and in investigating their properties~\cite{EGB1a,EGB1b,EGB1c,EGB1d,EGB2,EGB3}, rotating black holes~\cite{EGB4a,EGB4b}, black holes coupled with magnetic charge~\cite{EGB5a,EGB5b,EGB5c,EGB5d}, Born–Infeld black holes~\cite{EGB5e}, relativistic stars~\cite{EGB6,EGB7}, gravitational lensing~\cite{EGB8,EGB9}, radiation and accretion phenomena~\cite{EGB10,EGB11}, stability issues~\cite{EGB12a,EGB12b,EGB12c}, quasi-normal modes~\cite{EGB13a,EGB13b}, black hole shadows~\cite{EGB14a,EGB14b,EslamPanah:2020hoj}, to~thermodynamics and phase transitions~\cite{EGB16a,EGB16b,EGB16c,EGB16d,HosseiniMansoori:2020yfj}, amongst other topics. Nevertheless,  a~number of questions
~\cite{EGB17a,EGB17b,EGB17c,EGB17d,EGB18,Arrechea:2020evj,Gurses:2020rxb,Arrechea:2020gjw} have been raised on the adopted approaches in~\cite{EGB0,EGB19}, and~some remedies have also been suggested to overcome these \mbox{difficulties~\cite{EGB18,EGB20,EGB21,EGB22}}.
{In a number of papers~\cite{EGB18,EGB21,EGB22}, several authors have attempted to solve this problem by inserting additional scalar fields, however, in~Refs.~\cite{Aoki:2020lig,Aoki:2020iwm,Aoki:2020ila}, it has been shown that if a $D$-dimensional solution of EGB gravity satisfy specific conditions such as possessing a vanishing Weyl tensor of the spatial metric, the~$D\rightarrow 4$ limit of this solution is a solution of the well-defined 4D EGB theory. In~Ref.~\cite{Jafarzade:2021}, it has been shown that any static and spherically symmetric solution of EGB gravity with just one metric function (including the solution studied in present paper) satisfies the required conditions so that its $D\rightarrow 4$ limit is a solution of well-defined 4D EGB gravity.}

In fact, research in astrophysical black holes has recently received much interest, due to the breakthrough discovery of the reconstruction of the event-horizon-scale images of the supermassive black hole candidate in the centre of the giant elliptical galaxy M87 by the Event Horizon Telescope project~\cite{KAkiyama1,KAkiyama2}. This was the first direct  evidence of the  existence of black holes consistent with the prediction of Einstein’s general relativity. The~image shows that there is a dark part surrounded by a bright ring, which are called the black hole shadow and photon ring, respectively.  The~ shadow  and  photon  ring  are  caused  from  the  light  deflection,  or~ gravitational lensing by the black holes. Indeed, the~gravitational field near the black hole’s event horizon is so strong that it can affect light paths and causes spherical light rings. The~black hole image provides us  with  information  concerning  jets  and  matter  dynamics  around  black  holes.  Moreover,  the~black  hole  shadow  is  one  of  the  useful  tools  for  comparing  alternative  theories of gravity with  general relativity and provides us with information of the black hole parameters including the mass, charge and rotation.
In fact, the~possibility of probing fundamental physics using such a shadow was studied in the context of extra dimensions~\cite{Vagnozzi:2019apd} and non-linear electrodynamics~\cite{Allahyari:2019jqz}.

After the EHT announcement, a~large amount of research has been devoted to determining the shadows of a vast class of BH solutions and the confrontation with the extracted
information from the EHT black hole shadow image of M87* \cite{Davoudiasl:2019,Jusufi:2019,Konoplya:2019}. By~doing so, one can obtain the allowed regions of the parameters of different theories of gravity for which the obtained shadow is consistent with the observational data (see for instance Refs.~\cite{Allahyari:2019jqz,Jafarzade:2021umv}). In~fact, this method can be an observational and experimental test to understand which of the generalized gravity theories are more consistent with the experimental data. Another interesting phenomenon appearing while studying black holes is the accretion disk. In~the sub-mm band, EHT's $\sim$25 $\mu as$ images of M87 revealed the first evidence for the central depression and ring of light associated with the accretion flow and photon orbits close to the black hole~\cite{KAkiyama1}. The~matter flow from the star to a companion BH creates the accretion disc and the system starts to irradiate in the X-ray diapason due to strong friction of the matter near the ISCO. Although~one cannot observe a black hole or its event horizon, the~accretion disk can be seen, releasing heat and powerful X-rays and gamma rays out into the universe as they smash into each other. The~study of X-ray emission from accreting supermassive black holes have suggested that the primary source of X-rays is very compact and a large fraction of this emission is reflected and reprocessed by the innermost parts of the accretion disc. 
X-ray imaging can clearly reveal how far the accretion disc extends down towards the black hole. The~spacetime structure of the central black hole govern two basic classes of accretion disks: ($i$) the geometrically thin, Keplerian, accretion disks whose structure is mainly governed by the spacetime circular geodesics~\cite{Novikov:1973} and ($ii$) the geometrically thick, toroidal accretion disks governed by the effective potential of an orbiting perfect fluid determined by the Euler equation~\cite{Kozlowski:1978}. Most of the observed black hole candidates have an accretion disk constituted from conducting plasma whose dynamics can generate a magnetic field external to the black hole. In~both thin and thick accretion disks, magnetic fields could play very important roles. The~internal magnetic fields are crucial due to the magneto-rotational instability generating the disks' viscosity causing accretion itself, while the external magnetic fields could substantially modify the structure of the disk and create non-equatorial charged-particle circular orbits~\cite{Kovar:2010}. Motivated by specific interesting properties of the accretion disk, much research has been undertaken, including accretion disks surrounding a BH as a source of QPO in different models~\cite{Rezzolla:2003,Torok:1103,Stuchlik:2016}, studying the efficiency of energy release from the accretion disk~\cite{Rayimbaev:2021}, images of the Kaluza-Klein black hole surrounded with thin accretion disk~\cite{Mirzaev:2022} and influence of cosmic repulsion and magnetic fields on accretion disks~\cite{Stuchlik:2020}.

An interesting way to analyse the behaviour and stability of black holes is the study of their quasinormal modes (QNMs). QNMs are the response of black holes to external perturbations, and~they appear as damping oscillations after the initial outbursts of radiation. QNMs have been observed in a recent series of experiments by the LIGO/VIRGO collaborations when detecting gravitational waves from astrophysical black holes~\cite{BPAbbott1,BPAbbott2,BPAbbott3}. These frequencies depend on the black hole properties such as the mass, charge and the angular momentum and on the type of the perturbation (scalar, Dirac, vector or tensor), but~not on the initial conditions of the perturbations~\cite{Panotopoulos,RAKonoplya}. The~QN frequencies which are characterized by  complex numbers, $\omega = \omega_{R} - i\omega_{I}$, encode the information of how a black hole relaxes after a perturbation. The~sign of the imaginary part indicates if the mode is stable or unstable. For~$\omega_{I} < 0$ (exponential growth), the~mode is unstable, whereas for $\omega_{I} > 0$ (exponential decay) it is stable. In~the case of a stable mode, the~real part gives the frequency of the oscillation, $\omega_{R}/2\pi$, while the inverse of $|\omega_{I}|$ determines the damping time, $t^{-1}_{D} = |\omega_{I}|$ \cite{ARincon}.

There are several approaches to study the black hole’s QNMs including the WKB approximations, monodromy methods, series solutions in asymptotically AdS backgrounds and Leaver’s continued fraction method. Among~the above-mentioned  methods, the~WKB approach has received much interest during the past decades as it provides sufficient accuracy. This method was suggested by Schutz and Will at the first order~\cite{Schutz} and then later developed to higher orders~\cite{SIyer,Konoplya1,Matyjasek}. The~studies in the context of QNMs showed that there exists a close relation between them and phase transitions of black holes~\cite{Shen:2007xk,JJing1,XHeB}.
The phase transition is a thermodynamic phenomenon, while QNMs which are determined by the intrinsic properties of black holes are related to the dynamics. Such a connection enriches the relationship between the black hole dynamics and black hole thermodynamics. 
The relation between the quasinormal frequencies  and the thermodynamical quantities at the eikonal limit has been studied for static solutions~\cite{SHod} and rotating cases~\cite{SMusiri}. In~Ref.~\cite{VCardoso}, it was shown that the real and imaginary parts of the QNMs are related to the geodesic angular velocity and the Lyapunov exponent, respectively. Furthermore, a~connection between QNMs in the eikonal limit and lensing in the strong deflection limit has been obtained in Ref.~\cite{StefanovS}.
Recently, a~connection between the QNMs and the shadow radius has been found for static black holes~\cite{Jus1} and rotating solutions~\cite{Jus2}.

This paper is organized as follows. In~Section~\ref{secII}, we briefly review the novel 4D EGB gravity coupled to exponential nonlinear electrodynamics and present the AdS black hole solution, which we analyse throughout this work. In~Section~\ref{secIII}, we present a study of the optical features of the AdS black hole geometry. More specifically, in~Section~\ref{secIIIa}, we investigate the shadow behavior of this black solution and discuss the influence of the black hole parameters on the size of photon orbits and the spherical shadow. In~Section~\ref{secIIIb}, we calculate the energy emission rate and explore the effect of different parameters on the emission of particles around the black hole. In~Section \ref{secIIIc}, we further analyse the effective role of these parameters on the light deflection angle around this kind of black holes. In~Section \ref{secIIId}, we present a study of quasinormal modes of scalar perturbations and discuss the influence of the parameters. Finally, In Section~\ref{sec:conclusion}, we summarize the results and~conclude.


\section{Action and Black Hole~Solution}\label{secII}


The action of EGB gravity coupled to exponential nonlinear electrodynamics (ENED) is expressed as~\cite{DVSingh}
\begin{equation}
S=\frac{1}{16\pi}\int d^{D}x\sqrt{-g}\left[ R-2\Lambda+\alpha \mathcal{G}-2\mathcal{L}(F)\right],
\label{EqNaction}
\end{equation}
where the cosmological constant $ \Lambda $ is related to the AdS length as $\Lambda=-{(D-1)(D-2)}/{2l^2}$, $\alpha$ is the GB coupling constant and $\mathcal{G} $ is the GB quadratic curvature correction given by
\begin{equation}
\mathcal{G} =R^{2}-4R_{\mu\nu}R^{\mu\nu}+R_{\mu\nu\rho\sigma}R^{\mu\nu\rho\sigma}.
\end{equation}
The Lagrangian density $ \mathcal{L}(F) $ is a function of the invariant $F=\frac{1}{4} F_{\mu\nu}F^{\mu\nu} $, where \linebreak\mbox{$F_{\mu\nu} = \partial_{\mu}A_{\nu}-\partial_{\nu}A_{\mu}$} and $A_{\mu}$ is the potential. The~specific form used in this work is of the exponential type~\cite{Ghosh:2018bxg} defined as
\begin{eqnarray}
\mathcal{L}(F) =\beta\,F \exp \left[-k q^{-\gamma}(2F)^{\zeta}\right],
\end{eqnarray}
where $\beta={(D-2)(D-3)}/{2} $, $\gamma = {(D-3)}/{(D-2)} $, $\zeta={(D-3)}/{(2D-4)}$, and~in which $ q $ and $ k $ are the magnetic charge and the NED parameter, respectively~\cite{DVSingh}\endnote{We refer the reader to Ref.~\cite{Kruglov:2017fck} for an interesting solution of a black hole as a magnetic monopole within exponential nonlinear electrodynamics.}.

Varying the action (\ref{EqNaction}) results to the following equations of motion
\begin{eqnarray}
{G}_{\mu \nu} + \Lambda g_{\mu \nu} +\alpha {H}_{\mu \nu} = \mathcal{T}_{\mu \nu}\,,
 \label{FieldEq} 
\end{eqnarray}
and
\begin{eqnarray}
\nabla_{\mu}\left(\frac{\partial {\mathcal{L}(F)}}{\partial F}F^{\mu \nu}\right)=0\,, \qquad \nabla_{\mu}(^*F^{\mu \nu})=0,
\label{egb3}
\end{eqnarray}
respectively. The~Einstein tensor $G_{\mu \nu}$ and Lanczos tensor  $H_{\mu \nu} $ are defined by
\begin{eqnarray}
G_{\mu \nu}&=&R_{\mu \nu}-\frac{1}{2}g_{\mu \nu}R ,\nonumber\\
{H}_{\mu \nu}&=&2\left[RR_{\mu \nu}-2R_{\mu \sigma}R^{\sigma}_{\nu}-2R^{\sigma\lambda}R_{\mu \sigma \nu \lambda} +R_{\mu}^{~\sigma \lambda \rho}R_{\nu \sigma \lambda \rho}\right] -\frac{1}{2}g_{\mu\nu}{\mathcal{L}}_{GB}, \nonumber
\end{eqnarray}
and the energy-momentum tensor is given by
\begin{eqnarray}
\mathcal{T}^{\mu}_{\nu} \equiv 2\left[\frac{\partial {\mathcal{L}(F)}}{\partial F}F^{\mu\sigma}F_{\nu\sigma}-\delta^{\mu}_{\nu}{\mathcal{L}(F)}\right].
\end{eqnarray}

We consider a static and spherically symmetric $D$-dimensional metric ansatz
\begin{equation}
ds^2 = -f(r)dt^2+\frac{1}{f(r)} dr^2 + r^2 d\Omega^2_{D-2}.
\label{metric}
\end{equation}
In the limit $D\to 4$,  the~solution is given by~\cite{DVSingh,Ghosh:2018bxg}
\begin{equation}
f_{\pm}(r)=1+\frac{r^{2}}{2\alpha}\left[1\pm\sqrt{1+4\alpha\left(\frac{2M e^{-k/r}}{r^{3}}-\frac{1}{l^2}\right)} \right],
\label{Eqsol1}
\end{equation}
where  $ M $ can be identified as the mass of the black hole. In~the limit of $k = 0$, the~above form 
reduces to the Glavan and Lin solution~\cite{EGB0}. 
For the case $r\gg k$ the solution (\ref{Eqsol1}) takes the form
\begin{eqnarray}
f_{\pm}(r)&=&1+\frac{r^2}{2\alpha}\left[1 \pm  \sqrt{1+4\alpha\left(\frac{2M }{r^3} - \frac{q^2}{r^4}-\frac{1}{l^2}\right)}\,\right]
  + \mathcal{O}\left(\frac{1}{r^3}\right),  \nonumber
\label{sol2}
\end{eqnarray}
which is the form of a charged black hole in  4D EGB gravity with a negative cosmological constant~\cite{EGB2}, by~identifying the electric charge as $q^2 = 2Mk$ \cite{Ghosh:2018bxg}. 

Note that in the limit $\alpha \to 0$ or large $r$,  the~solution (\ref{Eqsol1})  behaves asymptotically as
\begin{eqnarray}
&&f_-(r) \approx 1-\frac{2M e^{-k/r}}{r} + \frac{r^2}{l^2} + \mathcal{O}\left(\frac{1}{r^3}\right),\nonumber\\
&&f_+({r})\approx 1+\frac{2M e^{-k/r}}{r} -\frac{r^2}{l^2}+\frac{r^2}{\alpha}+ \mathcal{O}\left(\frac{1}{r^3}\right), \nonumber
\end{eqnarray}
where the negative branch corresponds to a regular 4D AdS black hole, while the positive branch leads to instabilities of the graviton, due to the opposite sign in the mass~\cite{DVSingh}. 
Thus, since only the negative branch leads to a physically meaningful solution, we will limit our analysis to this branch of the~solution.  


\section{Optical Features of the AdS Black Hole~Spacetime}\label{secIII}

In this section, we present an in-depth study of the optical features of AdS black holes in the novel 4D EGB gravity coupled to ENED, given by the solution (\ref{Eqsol1}), such as the shadow geometrical shape, the~energy emission rate, the~deflection angle and quasinormal modes. Taking into account these dynamic quantities, we investigate the effects on the black hole solution by varying the parameters of the~theory. 

\subsection{Photon Sphere and~Shadow}\label{secIIIa}

Here, we are interested in investigating  the shadow of the black hole solution (\ref{Eqsol1}), with~the negative branch, and~exploring the effect of the GB coupling constant, the~NED parameter and the cosmological constant on the radius of the photon sphere and the spherical shadow. To~do so, we take into account the Hamilton-Jacobi method for null curves in the black hole spacetime as~\cite{Carter,M.Zhang}
\begin{equation}
\frac{\partial S}{\partial \sigma}+H=0,
\end{equation}
where $S $ and $\sigma $ are the Jacobi action and affine parameter along
the geodesics, respectively. 

A photon travelling along null geodesics in a static and spherically symmetric spacetime is governed by the following Hamiltonian: $H=\frac{1}{2}g^{ij}p_{i}p_{j}=0$.
%
Since the black hole solution (\ref{Eqsol1}) is spherically symmetric, we 
consider photons moving on the equatorial plane with $\theta=\pi/2$, without~a significant loss of generality. Thus, the~Hamiltonian can be written~as
\begin{equation}
\frac{1}{2}\left[-\frac{p_{t}^{2}}{f(r)}+f(r)p_{r}^{2}+\frac{p_{\phi}^{2}}{r^{2}}\right] =0
\label{EqNHa}
\end{equation}

Using the fact that the Hamiltonian is independent of the coordinates $t$ and $\phi$, one can define two constants of motion as
\begin{equation}
p_{t}\equiv\frac{\partial H}{\partial\dot{t}}=-E , \qquad  p_{\phi}\equiv\frac{\partial H}{\partial%
\dot{\phi}}=L ,  \label{Eqenergy}
\end{equation}
where the quantities $E$ and $L$ are the energy and  angular momentum of the photon, respectively. Using the Hamiltonian formalism, the~equations of
motion can be derived~as
\begin{eqnarray}
\dot{t}&=&\frac{\partial H}{\partial p_{t}}=-\frac{p_{t}}{f(r)}, \qquad
\dot{r}=\frac{\partial H}{\partial p_{r}}=p_{r}f(r), \qquad
\dot{\phi}=\frac{\partial H}{\partial p_{\phi}}=\frac{p_{\phi}}{r^{2}}.  
	\nonumber
\end{eqnarray}
%
where $p_r$ is the radial momentum and the overdot denotes a derivative with respect to the affine parameter $\sigma $. These equations provide a complete description of the dynamics by taking into account the orbital equation of motion, $\dot{r}^{2}+V_{\rm eff}(r)=0$, where the effective potential is defined as
\begin{equation}
V_{\rm eff}(r)=f(r)\left[ \frac{ L^{2}}{r^{2}}-\frac{E^{2}}{f(r)}\right] .  \label{Eqpotential}
\end{equation}

Since the photon orbits are circular and unstable associated to the maximum value of the effective potential, we use  the following conditions to obtain such a maximum value,
\begin{equation}
V_{\rm eff}(r) \big \vert_{r=r_{p}}=0,  \qquad \frac{\partial V_{\rm eff}(r)}{\partial r}%
\Big\vert_{r=r_{p}}=0,  \label{Eqcondition}
\end{equation}
which result in the following equation
\begin{eqnarray}
(3+4\alpha\Lambda)r_{p}^{4}-3M^{2} e^{-2k/r_{p}} (3r_{p}-k)^{2}
+24 \alpha M r_{p} e^{-k/r_{p}}=0.
\label{EqShadowN}
\end{eqnarray}

Equation~(\ref{EqShadowN}) is complicated to solve analytically, so we employ numerical methods to obtain the radius of the photon sphere. To~this effect, several values of the event horizon $ (r_{e}) $ and  photon sphere radius $ (r_{p}) $ are presented in Table~\ref{table3}. We verify that increasing values of the NED and GB parameters tend to decrease $r_{e} $ and $r_{p} $. Thus, specific constraints should be imposed on these parameters in order to have a real event horizon. Regarding the effect of the cosmological constant, we verify that as $ \Lambda $ increases from $ -0.17 $ to $ -0.01 $, the~event horizon (the radius of photon sphere) increases (decreases).  

 \begin{table}[H]
\caption{Event horizon and photon sphere radius for variation of the cosmological constant, the~NED and the GB parameters for the specific case of $M =1$.} \label{table3}
\small
\setlength{\tabcolsep}{4.42mm}
\begin{tabular}{c c c c c c c}
\toprule
$k$   & $0$  &
 $0.1$ &  $0.3$ & $0.5$ & $ 0.6 $  & \\ \midrule
$ r_e (\alpha=0.2 $, $\Lambda=-0.02 $) & $ 1.84 $ & $ 1.73 $ & $1.47$ & $1.02$& 0.8--0.3 $I$ \\ 
$ r_p (\alpha=0.2 $, $\Lambda=-0.02 $) & $ 2.91 $ & $ 2.76 $ & $2.43$ & $1.99$&$1.62$\\\midrule
$\alpha$   & $0.02$  &
 $0.2$ &  $0.4$ & $0.6$ & $ 0.7 $ \\ \midrule
$ r_e (k=0.2 $, $\Lambda=-0.02 $) & $ 1.73 $ & $1.61$ & $1.44$ & $1.14$& 0.9--0.2 $I$\\ 
$ r_p (k=0.2 $, $\Lambda=-0.02 $) & $ 2.70 $ & $2.60$ & $2.47$ & $2.30$&$ 2.19 $\\ \midrule
$\Lambda$   & $-0.01$  &
 $-0.05$ &  $-0.09$ & $-0.13$ &  $ -0.17 $  & \\ \midrule
$ r_e$ ($\alpha=k=0.2$) & $1.63$ &  $1.56$ & $1.51$ & $1.47$&$1.43$ \\
$ r_p$ ($\alpha=k=0.2$) & $2.60$ & $2.61$ & $2.63$ & $2.65$&$2.66$\\
 \bottomrule
\end{tabular}
\end{table}



We next analyse the radius of the shadow and explore the impact of the parameters of the theory. To~this effect, consider the orbit equation for the photon given by
\begin{equation}
\frac{dr}{d\phi}=\frac{\dot{r}}{\dot{\phi}}=\frac{r^{2}f(r)}{L}p_{r}.  \label{Eqorbit}
\end{equation}
Using Equation~(\ref{EqNHa}) and the constraint $dr/d\phi\vert_{r=R}=0 $, one finds
\begin{equation}
\frac{dr}{d\phi}=\pm r\sqrt{f(r)\left[\frac{r^{2}f(R)}{R^{2}f(r)} -1\right] }.  \label{EqTp}
\end{equation}

Consider a light ray sent from a static observer at position $r_{0} $ and transmitted with an angle $\vartheta$ with respect to the radial direction, which results in~\cite{M.Zhang,Belhaj}
\begin{equation}
\cot \vartheta =\frac{\sqrt{g_{rr}}}{g_{\phi\phi}}\frac{dr}{d\phi}\Big\vert%
_{r=r_{0}}.  \label{Eqangle}
\end{equation}
Hence, the~shadow radius of the black hole as observed by a static observer at the position $r_0$ is given by
\begin{equation}
r_{s}=r_{0}\sin \vartheta =R\sqrt{\frac{f(r_{0})}{f(R)}}\Bigg\vert_{R=r_{p}}.
\label{Eqshadow}
\end{equation}

\begin{figure}[H]
\begin{adjustwidth}{-\extralength}{0cm}
\centering%
\subfigure[]
{\label{Fig7a}\includegraphics[width=.4\textwidth]{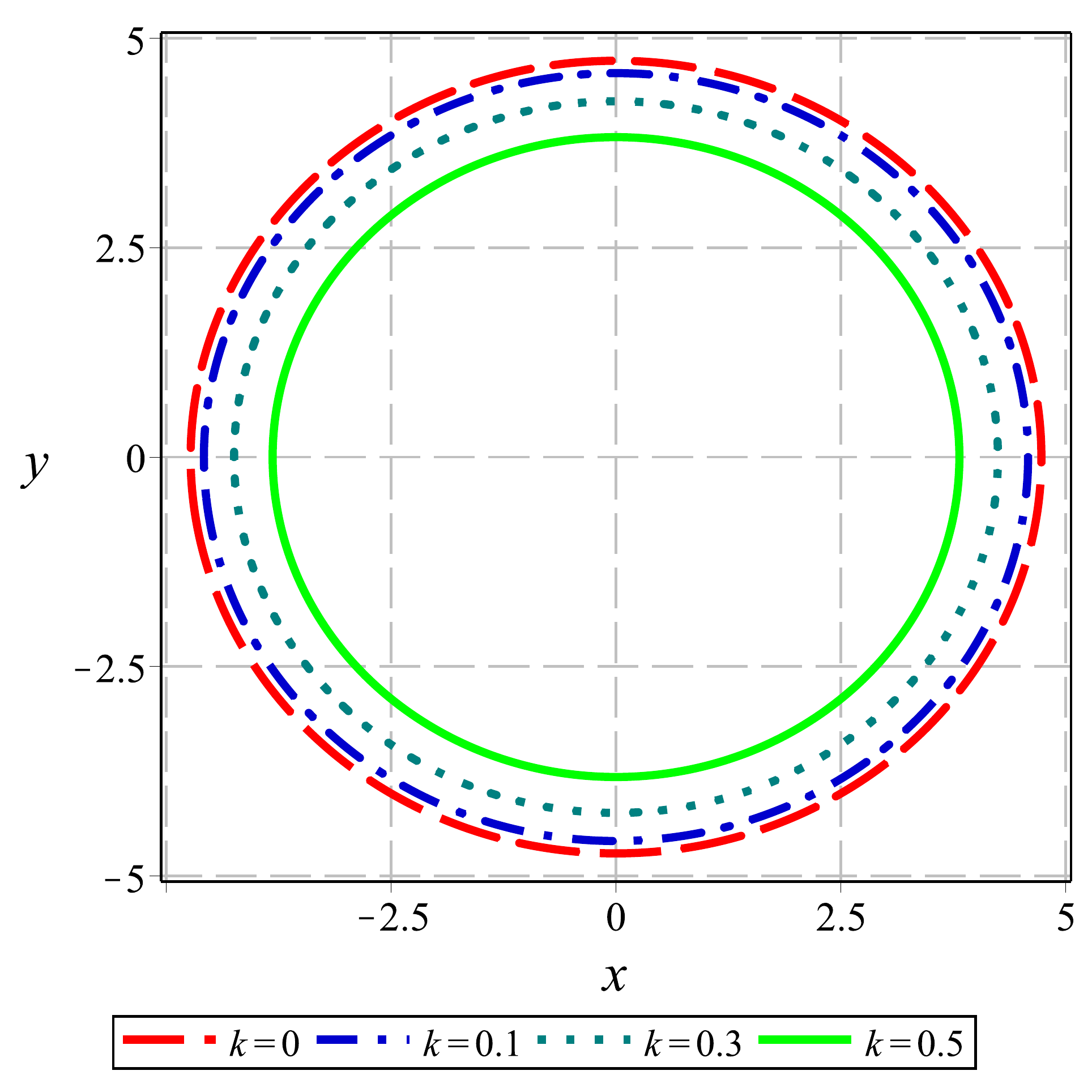}}
\subfigure[]
{\label{Fig7b}\includegraphics[width=.4\textwidth]{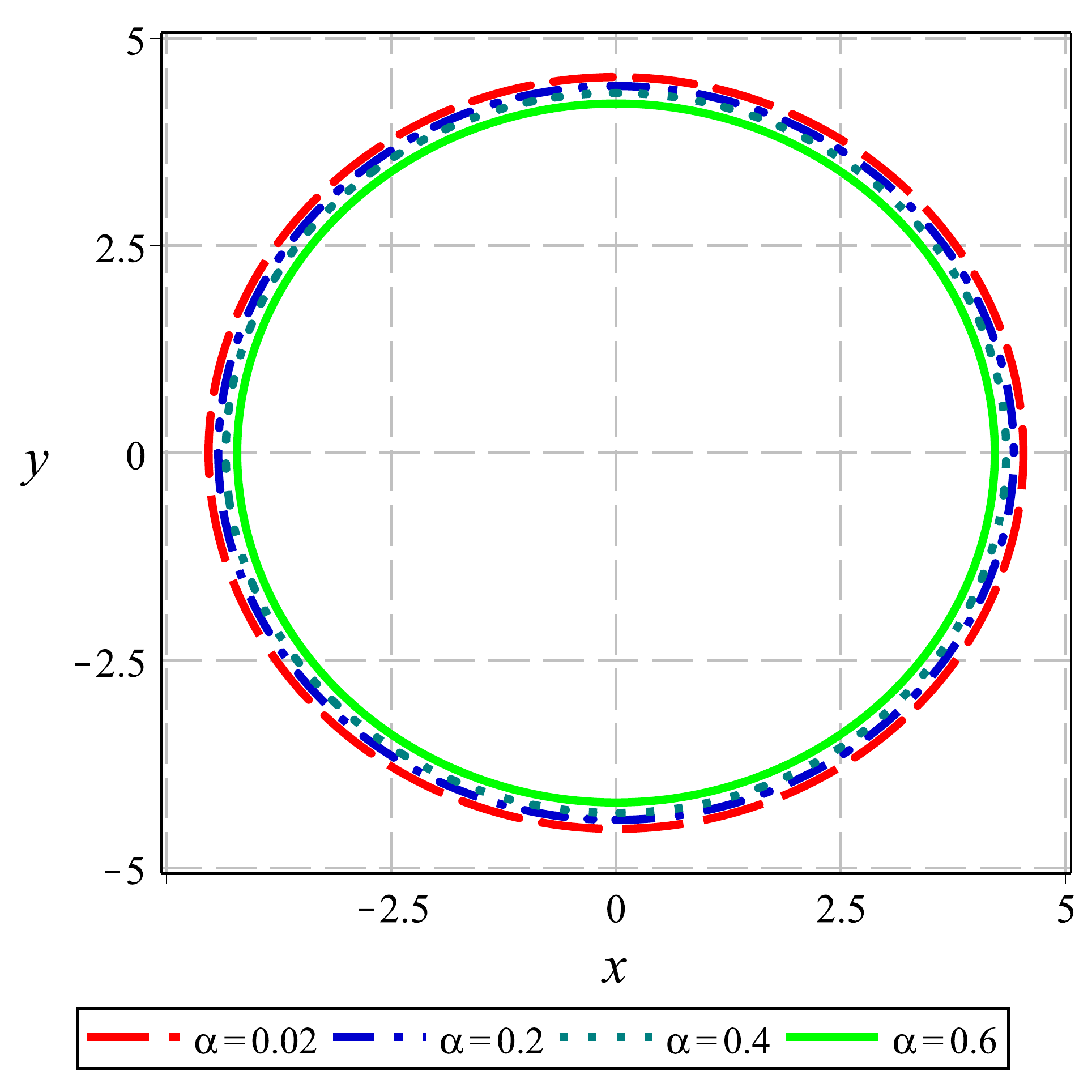}}
\subfigure[]
{\label{Fig7c}\includegraphics[width=.4\textwidth]{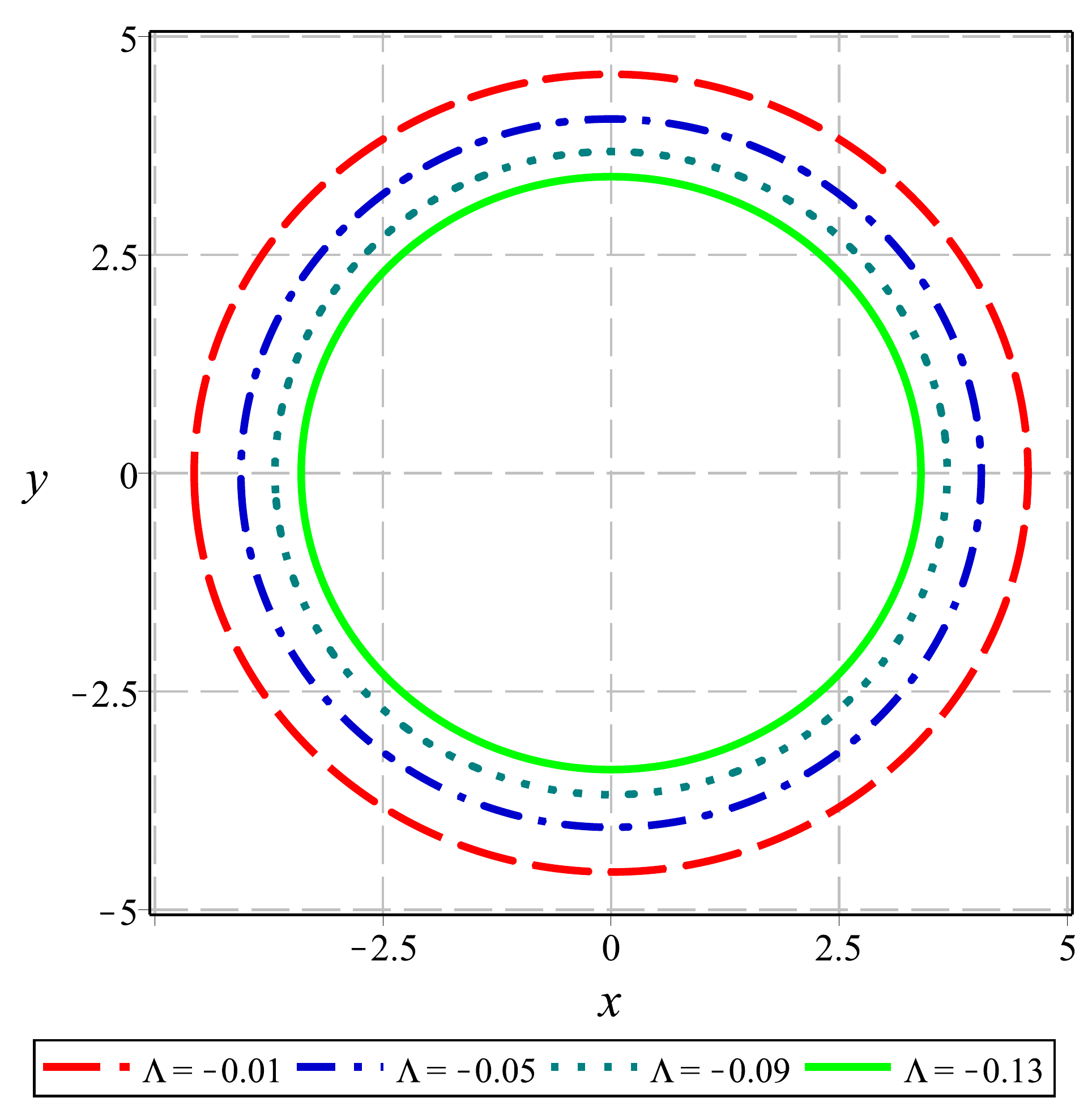}}
\end{adjustwidth}
\vspace{-6pt}
\caption{Black hole shadow in the Celestial plane $(x -y )$ for $M=1$.
Note that increasing the NED ($k$) and GB ($\alpha$) parameters, decreases the spherical shadow radius, as~depicted in Figure~\ref{Fig7}a,b, respectively, whereas increasing the cosmological constant, increases the radius (Figure~\ref{Fig7}c). See the text for more details. (\textbf{a}) $\alpha=0.2 $ and  $\Lambda=-0.02 $; (\textbf{b})~$k=0.2 $ and  $\Lambda=-0.02 $; (\textbf{c}) $k=\alpha=0.2$.}
\label{Fig7}
\end{figure}

The apparent shape of a shadow can be obtained by a stereographic projection in terms of the celestial coordinates $x$ and $y$ which are defined as
~\cite{Vazquez,RShaikh}
\begin{eqnarray}
x &=& \lim_{r_{0}\longrightarrow \infty} \left( -r_{0}^{2}\sin
\theta_{0}\frac{d\phi}{dr}\Big\vert_{(r_{0},\theta_{0})}\right), 
\end{eqnarray}
\begin{eqnarray}
y &=& \lim_{r_{0}\longrightarrow \infty} \left( r_{0}^{2}\frac{d\theta}{%
dr}\Big\vert_{(r_{0},\theta_{0})}\right).
\end{eqnarray}
where $( r_{0},\theta_{0}) $ are the position coordinates of the observer. 
Figure~\ref{Fig7} depicts the influence of the GB coupling constant, $\alpha$, the~NED parameter, $k$, and~the cosmological constant, $\Lambda$, on~the spherical shadow radius.
We notice that increasing the NED and GB parameters decreases the shadow radius, whereas increasing the cosmological constant consequently increases the shadow radius. Comparing Figure~\ref{Fig7}b with Figure~\ref{Fig7}a,c, we find that the variation of $ \alpha $ has a weaker effect on the shadow size than the other two parameters and a more significant effect occurs for the cosmological~constant.

\vspace{-12pt}

\subsection{Energy Emission~Rate}\label{secIIIb}



In this subsection, we study the effect of the parameters of the theory on the emission of particles around the black hole given by solution (\ref{Eqsol1}). It was shown that the black hole shadow corresponds to its high energy absorption cross section for the observer located at infinity~\cite{WWei,Belhaj1,Belhaj2}. In~fact,  at~very high energies, the~absorption cross-section oscillates around a limiting constant value $ \sigma_{\rm lim} $ which is approximately equal to the area of the photon sphere ($ \sigma_{\rm lim}\approx \pi r_{s}^{2} $).  The~energy emission rate is expressed as
\begin{equation}
\frac{d^{2}E(\omega)}{dtd\omega}=\frac{2\pi^{3}\omega^{3}r_{s}^{2}}{e^{\frac{\omega}{T}} -1},
\label{Eqemission}
\end{equation}
where $ \omega $ is the emission frequency and $T$ denotes the Hawking temperature, which for the present case is given by
\begin{equation}
T=\frac{-\Lambda r_{e}^{4}(3r_{e}-k)+3r_{e}^{2}(r_{e}-k)-3\alpha (r_{e}+k)}{12\pi r_{e}^{2}(r_{e}^{2}+2\alpha)}.
\end{equation}

\begin{figure}[H]
\begin{adjustwidth}{-\extralength}{0cm}
\centering%
\subfigure[]
{\label{Fig8a}\includegraphics[width=.4\textwidth]{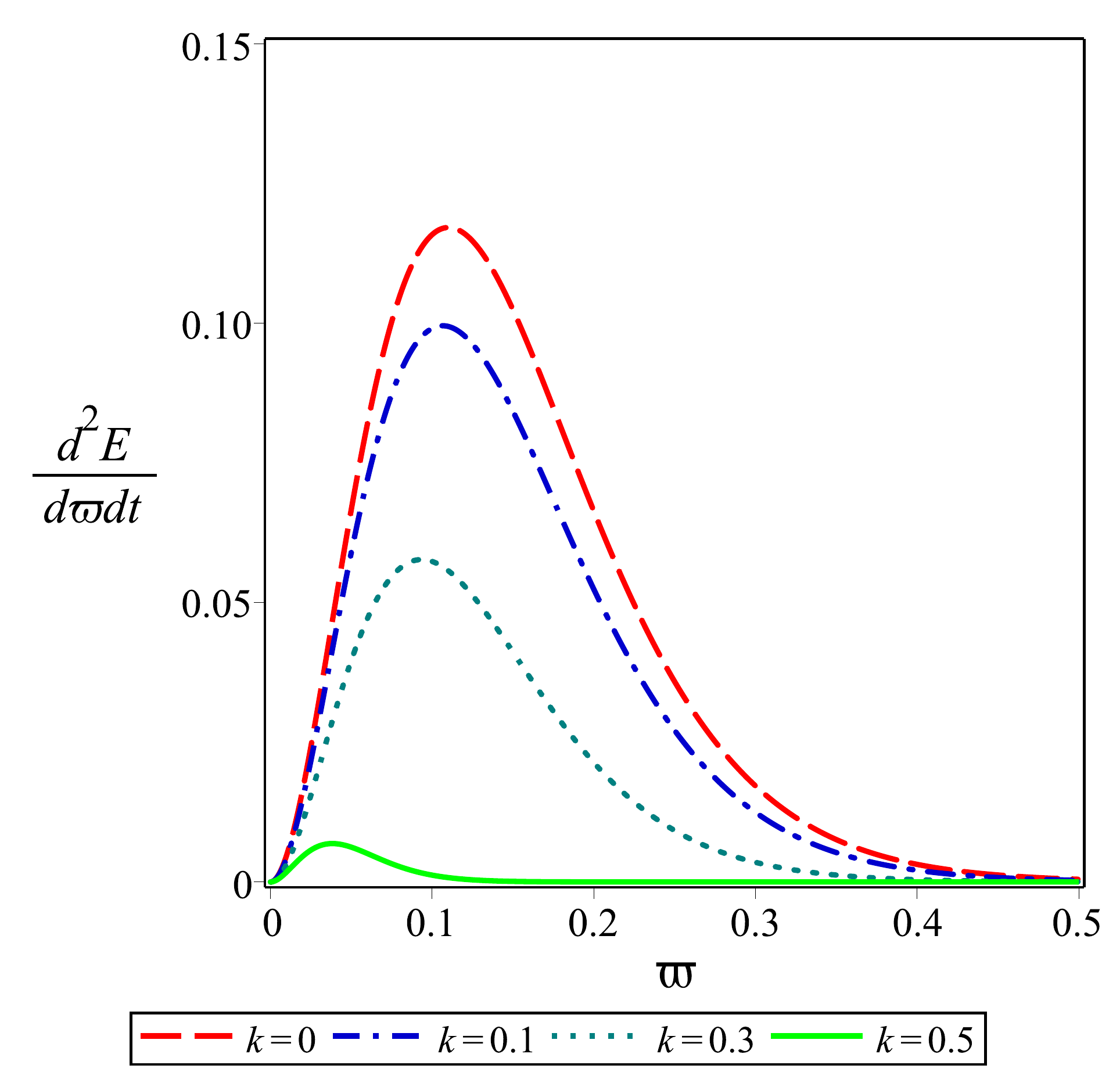} }
\subfigure[]
{\label{Fig8b}\includegraphics[width=.4\textwidth]{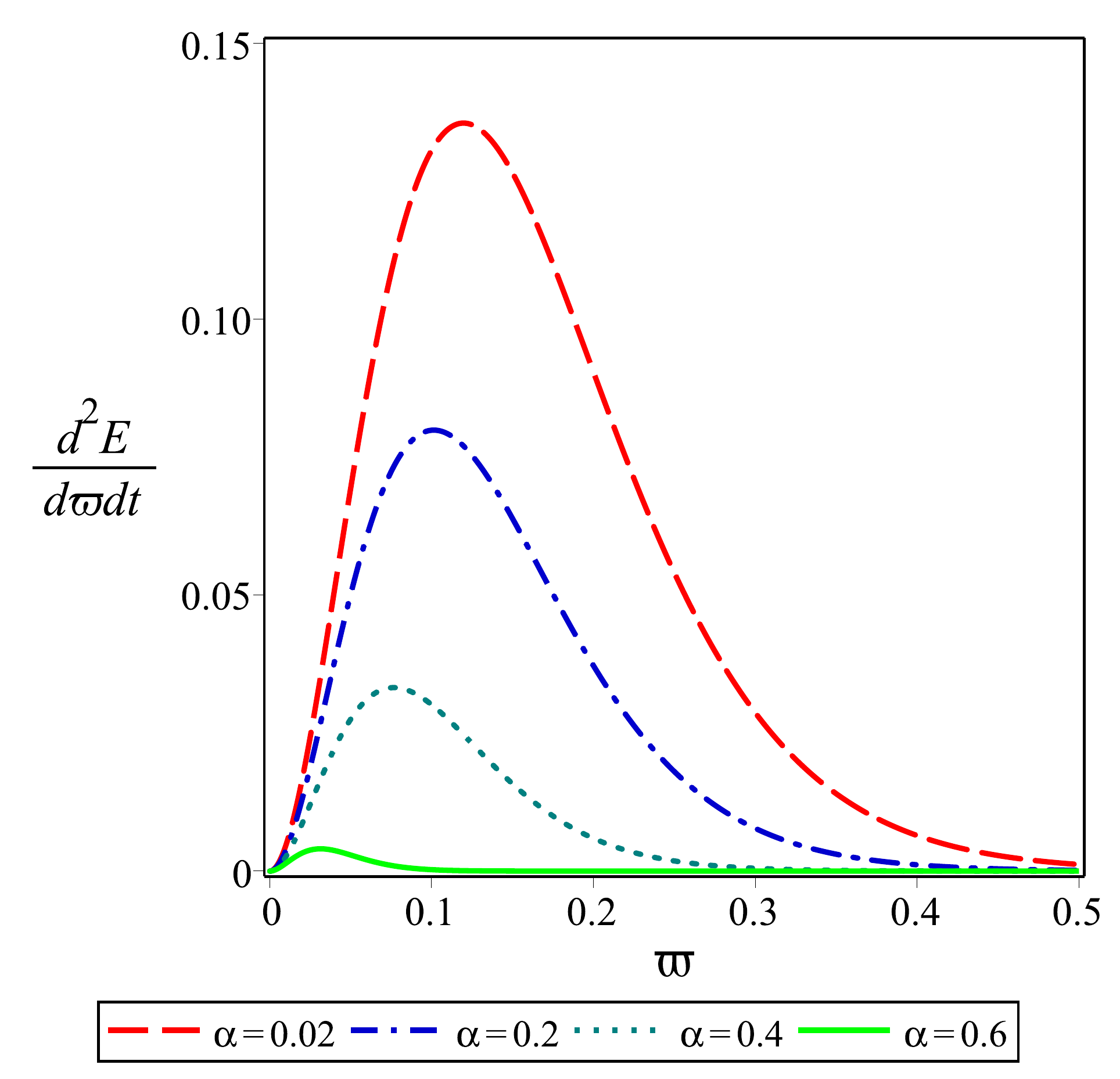}}
\subfigure[]
{\label{Fig8c}\includegraphics[width=.4\textwidth]{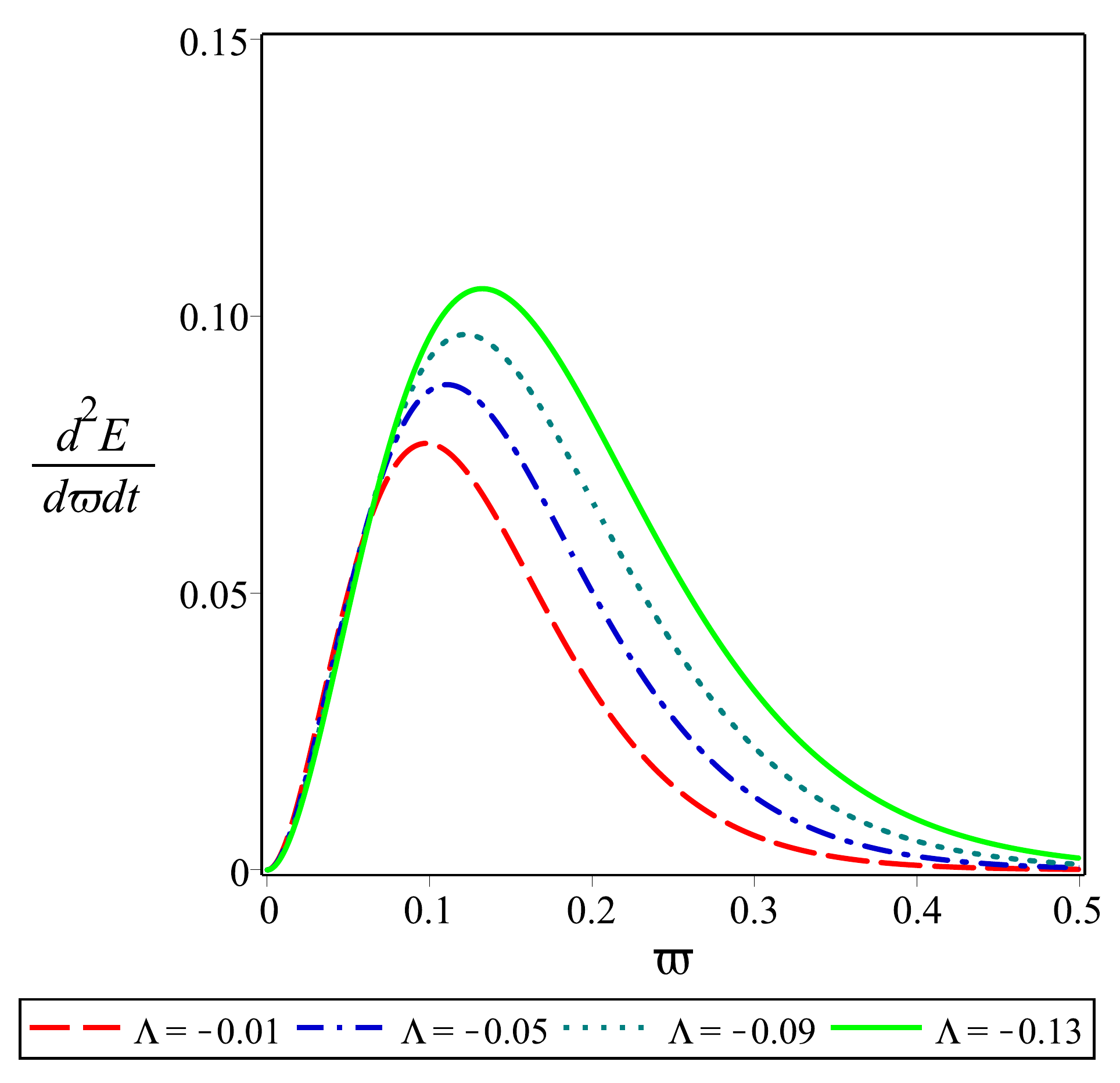} }
\end{adjustwidth}
\vspace{-6pt}
\caption{Energy emission rate of the black hole for $ M=1 $ and different values of $ k $, $ \alpha $ and $\Lambda$. It is transparent from the plots that the energy emission decreases with increasing values of these three parameters, and~that the radiation rate is highly affected by the GB coupling constant, whereas the variation of $ \Lambda $ has a weaker effect on the radiation rate compared to other two parameters. See the text for more~details. (\textbf{a}) $\alpha=0.2 $ and  $\Lambda=-0.02 $; (\textbf{b}) $k=0.2 $ and  $\Lambda=-0.02 $; (\textbf{c}) $k=\alpha=0.2$.}
\label{Fig8}
\end{figure}

In order to study the radiation rate of the solution (\ref{Eqsol1}), we obtain the energy emission rate from Equation~(\ref{Eqemission}). In~Figure~\ref{Fig8}, these energetic aspects are plotted as a function of the emission frequency  for different values of $ k $ (Figure~\ref{Fig8}a), $ \alpha $ (Figure~\ref{Fig8}b) and $ \Lambda $ (\mbox{Figure~\ref{Fig8}c}), respectively. It is transparent from the plots that the energy emission decreases with increasing values of these three parameters. This reveals the fact that as the coupling constants get stronger or when the curvature background becomes higher, the~energy emission rate becomes insignificant. In~fact, the~black hole has a longer lifetime in a high curvature background or with a stronger coupling. Comparing Figure~\ref{Fig8}b with Figure~\ref{Fig8}a,c, one verifies that the radiation rate is highly affected by the GB coupling constant, whereas the variation of $ \Lambda $ has a weaker effect on the radiation rate compared to the other two~parameters.

\vspace{-12pt}
\unskip

\subsection{Deflection~Angle}\label{secIIIc}



In this subsection, we explore the light deflection around the black hole solution (\ref{Eqsol1}). We employ the null geodesics method~\cite{Chandrasekhar,Weinberg,Kocherlakota,WJaved} to calculate the total deflection $\Theta$. This optical quantity can be obtained by the following relation
\begin{equation}
\Theta=2\int_{b}^{\infty}\Big\vert\frac{d\phi}{dr}\Big\vert dr -\pi
,  \label{EqDAn}
\end{equation}
where $b$ is the impact parameter, defined as $ b\equiv L/E$. Using Equation~(\ref{EqTp}), the~deflection angle is given by
\begin{eqnarray}
\Theta=\frac{2M e^{-k/b} \left(3-2 \alpha \Lambda \right)}{3k}\left(1+\frac{3b}{k}+\frac{6b^{2}}{k^{2}}+\frac{6b^{3}}{k^{3}}\right) \notag \\
-\frac{\alpha M^{2} e^{-2k/b} \chi}{k}+\frac{7}{3}+\frac{1}{9}\Lambda b^{2} \left(-3+ \alpha \Lambda \right),~
\end{eqnarray}
where
\begin{equation*}
\chi=\frac{2}{b^{3}}+\frac{6}{kb^{2}}+\frac{15}{bk^{2}}+\frac{30}{k^{3}}+\frac{45b}{k^{4}}+\frac{45b^{2}}{k^{5}}+\frac{45b^{3}}{2k^{6}} \,.
\end{equation*}

\begin{figure}[H]
\begin{adjustwidth}{-\extralength}{0cm}
\centering%
\subfigure[]
{\label{Fig9a}\includegraphics[width=0.4\textwidth]{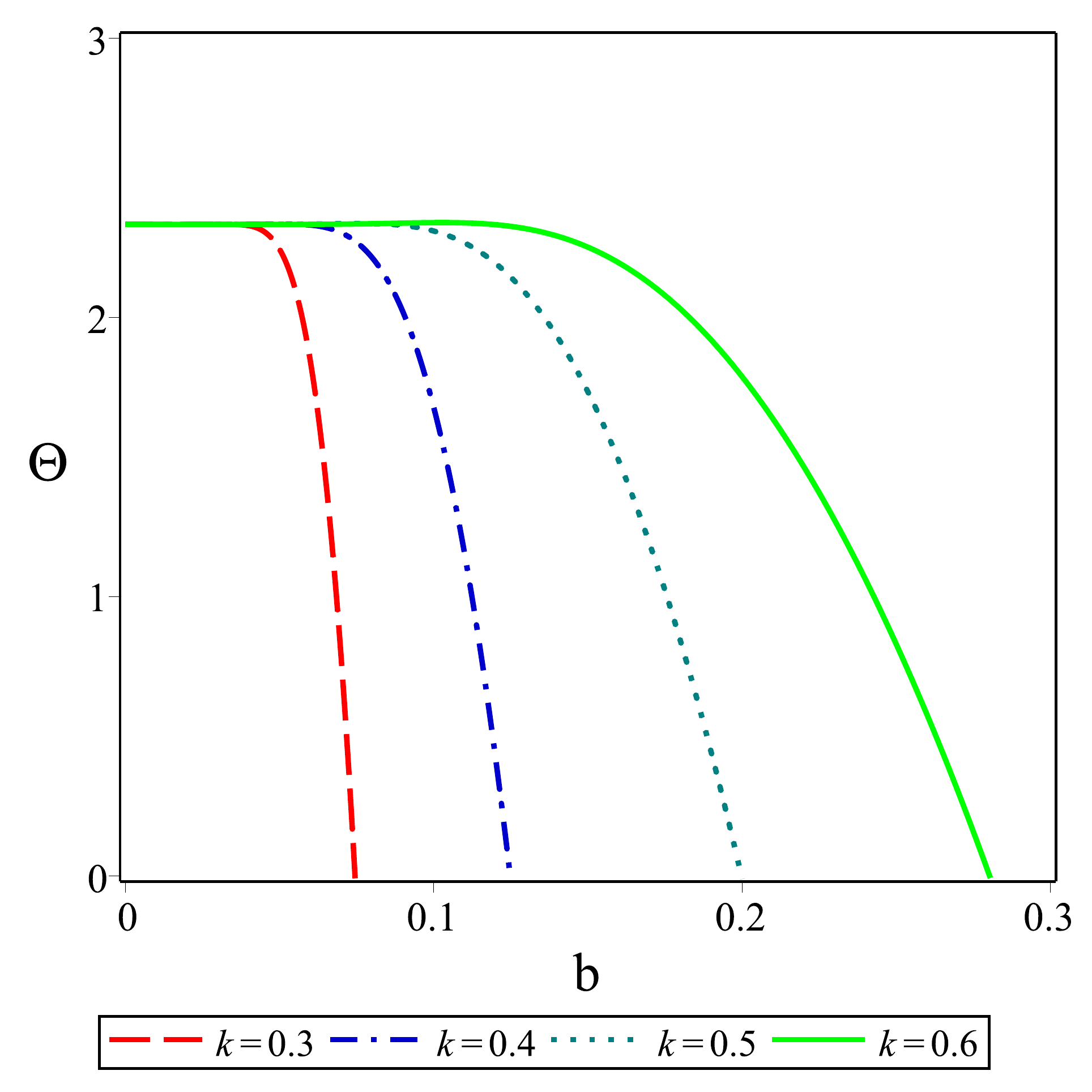}}
\subfigure[]
{\label{Fig9b}\includegraphics[width=0.4\textwidth]{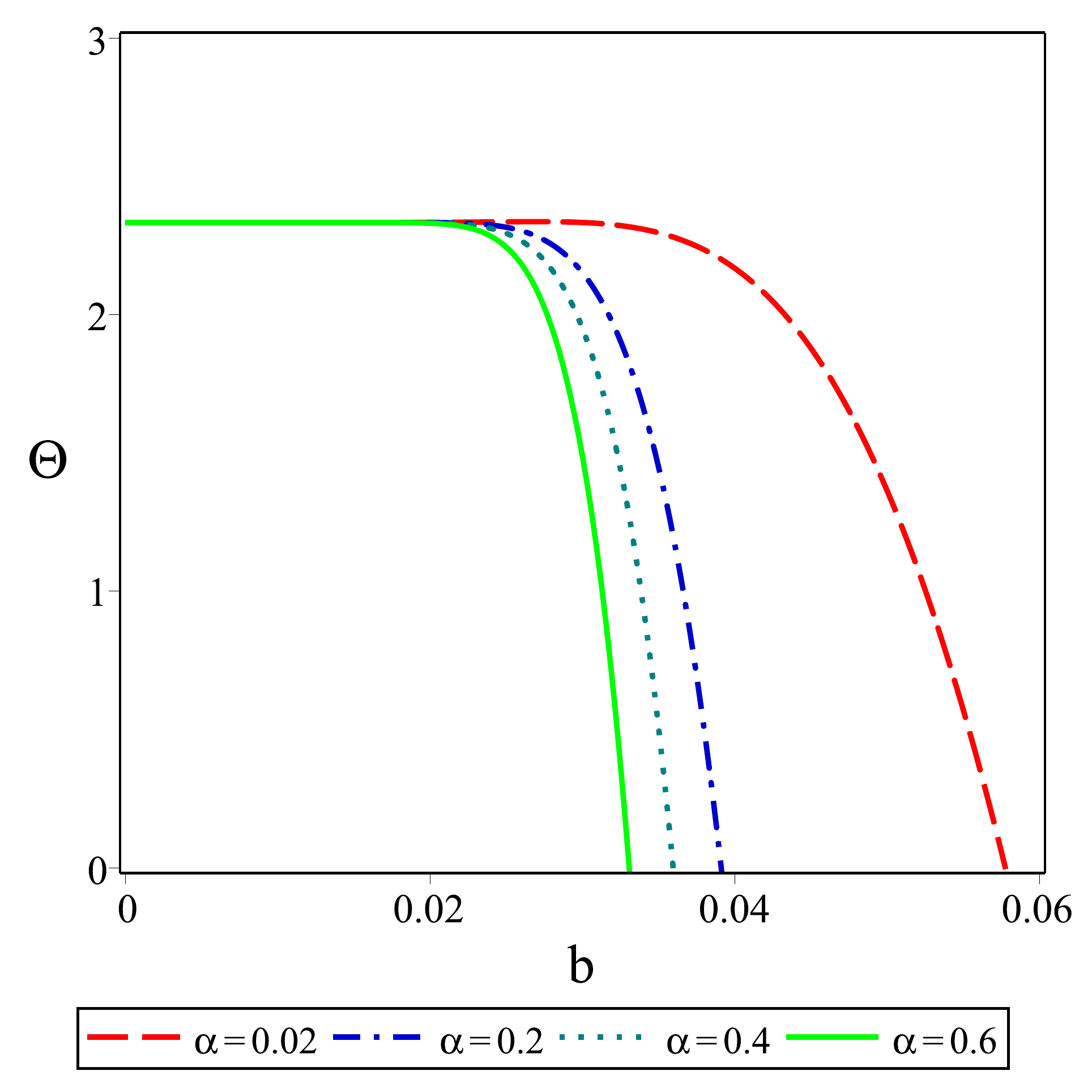}}
\subfigure[]
{\label{Fig9c}\includegraphics[width=0.4\textwidth]{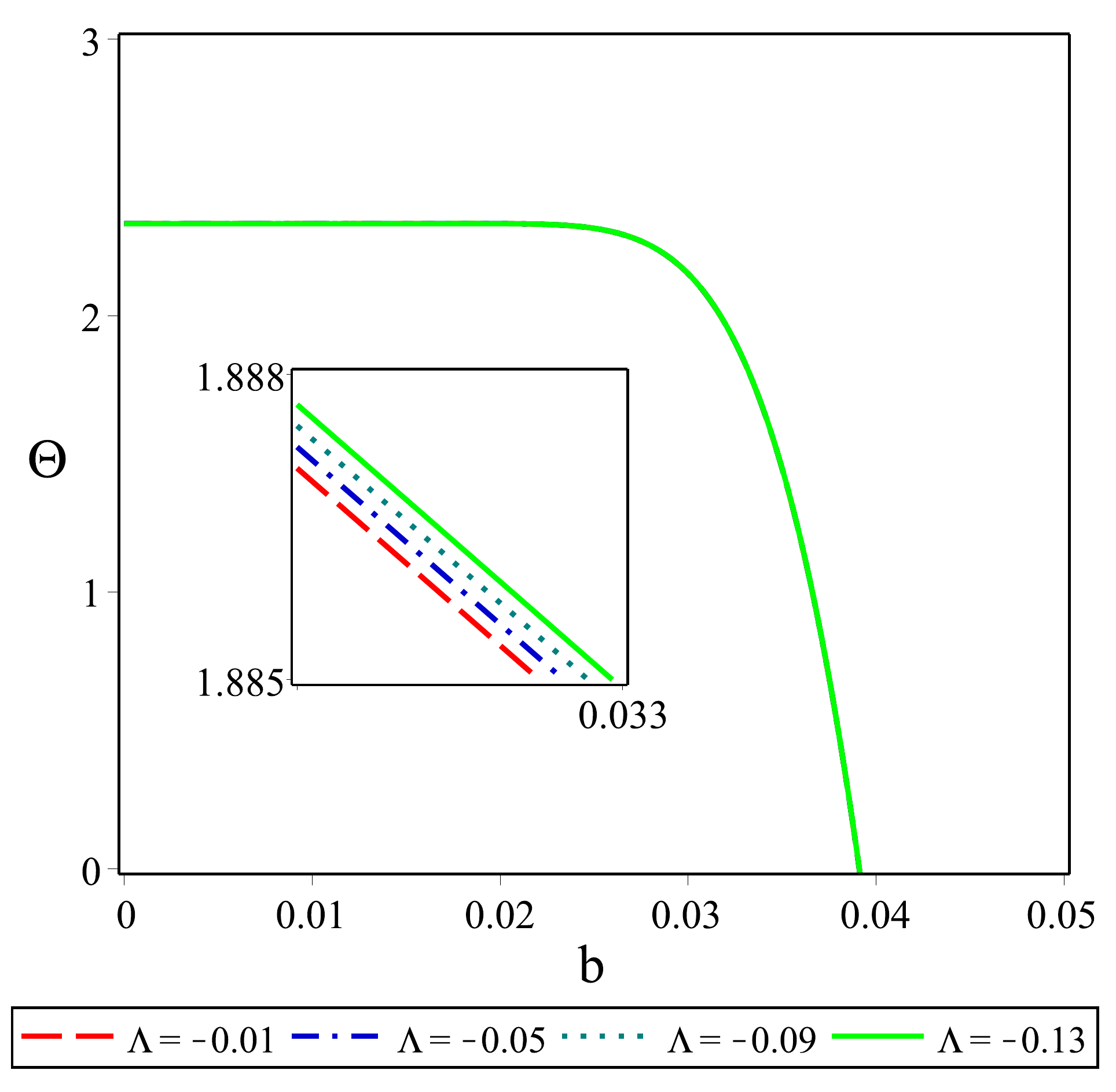}}
\end{adjustwidth}
\vspace{-6pt}
\caption{The behavior of $ \Theta $ with respect to the impact parameter $b$ for $ M=1 $ and different values of $ k $, $ \alpha $ and $  \Lambda$.
One verifies from Figure~\ref{Fig9}a, that the $ k $ parameter increases the light deflection, and~from Figure~\ref{Fig9}b that the GB parameter has an opposite effect on $ \Theta $. In~Figure~\ref{Fig9}c, we verify that the cosmological constant exhibits a decreasing contribution on the light deflection. See the text for more details. (\textbf{a})~$\alpha=0.2 $ and  $\Lambda=-0.02 $; (\textbf{b})~$k=0.2$ and  $\Lambda=-0.02$; (\textbf{c})~$k=\alpha=0.2$.}
\label{Fig9}
\end{figure}

The behavior of $ \Theta $ with respect to the impact parameter $ b $, is illustrated in Figure~\ref{Fig9}. As~we see, the~deflection angle is a decreasing function of $ b $. According to the relation $b\equiv L/E$, one can say that the light deflection is low when the energy of the photon decreases as compared to its angular momentum. This figure also demonstrates how the cosmological constant, the~NED and the GB parameters affect the deflection angle. From~Figure~\ref{Fig9}a, one verifies that the $ k $ parameter increases the light deflection, whereas, the~GB parameter has an opposite effect on $ \Theta $ (see Figure~\ref{Fig9}b). In~other words, the~effect of $  k$ and $ \alpha $ on the deflection angle is opposed to each other. In~Figure~\ref{Fig9}c, we investigate the influence of the cosmological constant on $ \Theta $ which displays its decreasing contribution on the light deflection. As~we see, this effect is negligible compared to the other two parameters. As~a result, one may conclude that the light deflection is low around this black hole solution with a stronger GB coupling or if it is located in a higher curvature~background.

\vspace{-12pt}
\unskip

\subsection{Quasinormal~Modes}\label{secIIId}





\begin{figure}[H]
\begin{adjustwidth}{-\extralength}{0cm}
\centering%
\subfigure[]
{\label{FigNVa} \includegraphics[width=0.5\textwidth]{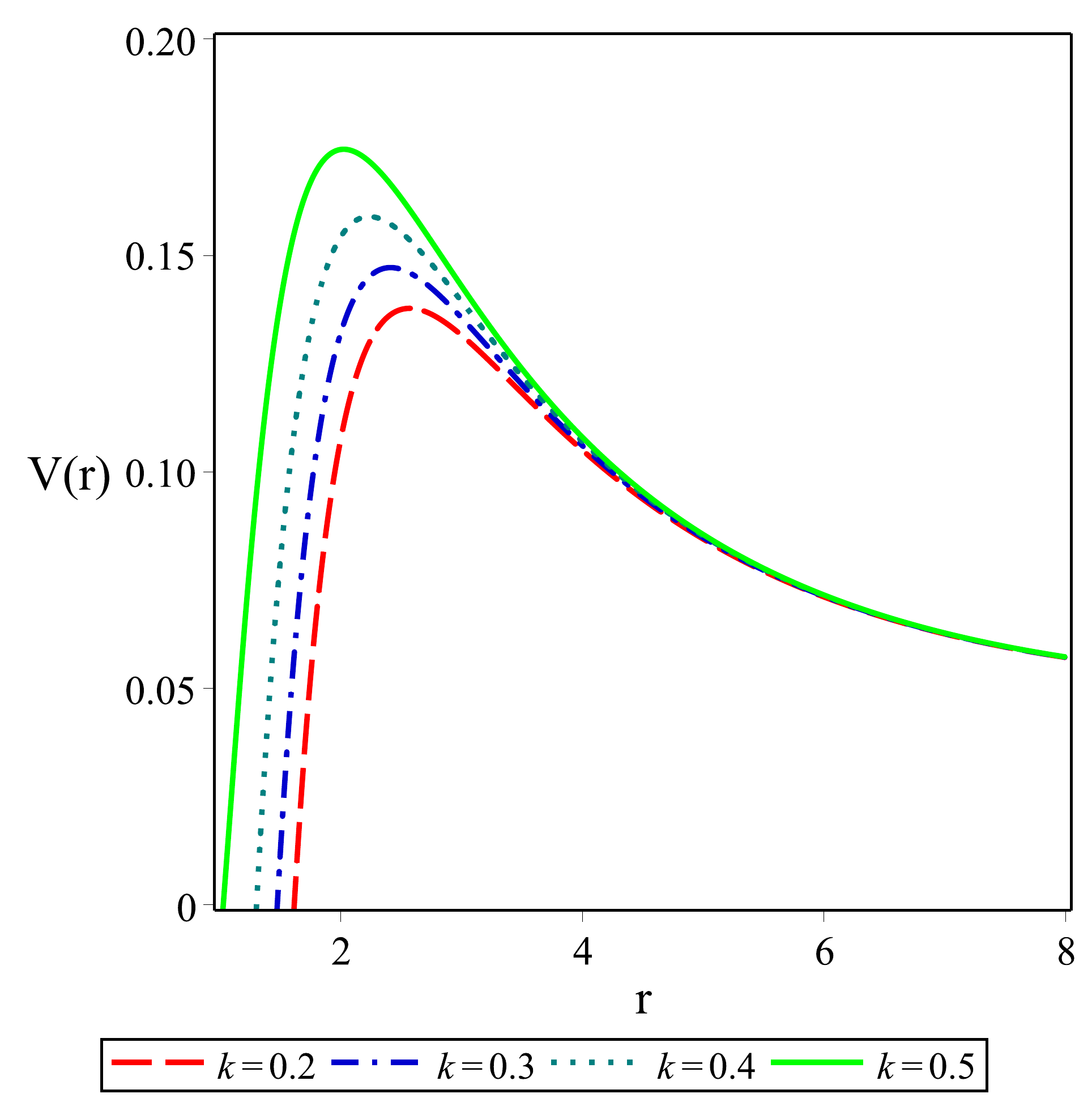}}
		\hspace{1.25cm}
\subfigure[]{
 \label{FigNVb}       \includegraphics[width=0.51\textwidth]{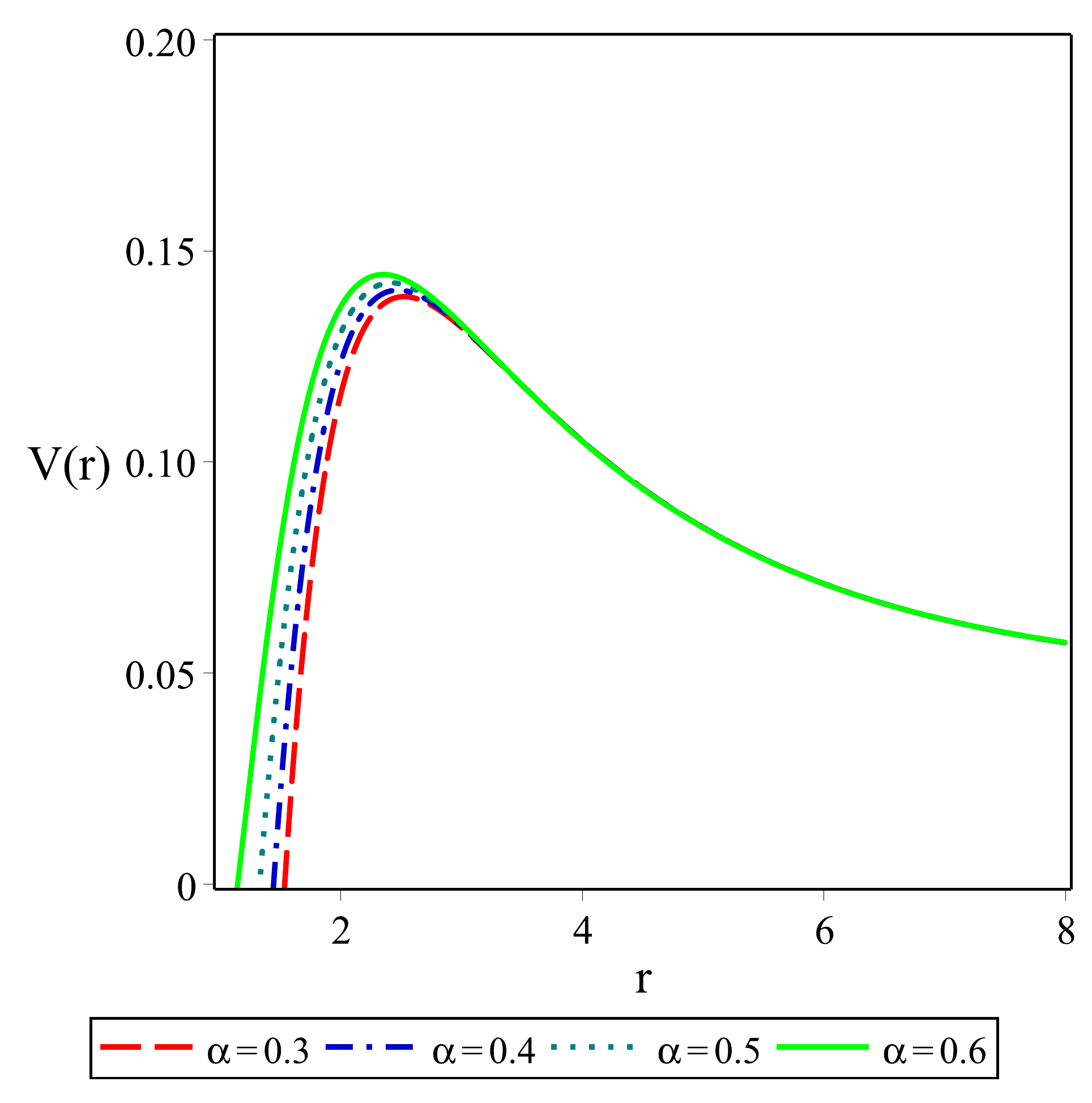}}
\\     
\subfigure[]{
       \label{FigNVc} \includegraphics[width=0.5\textwidth]{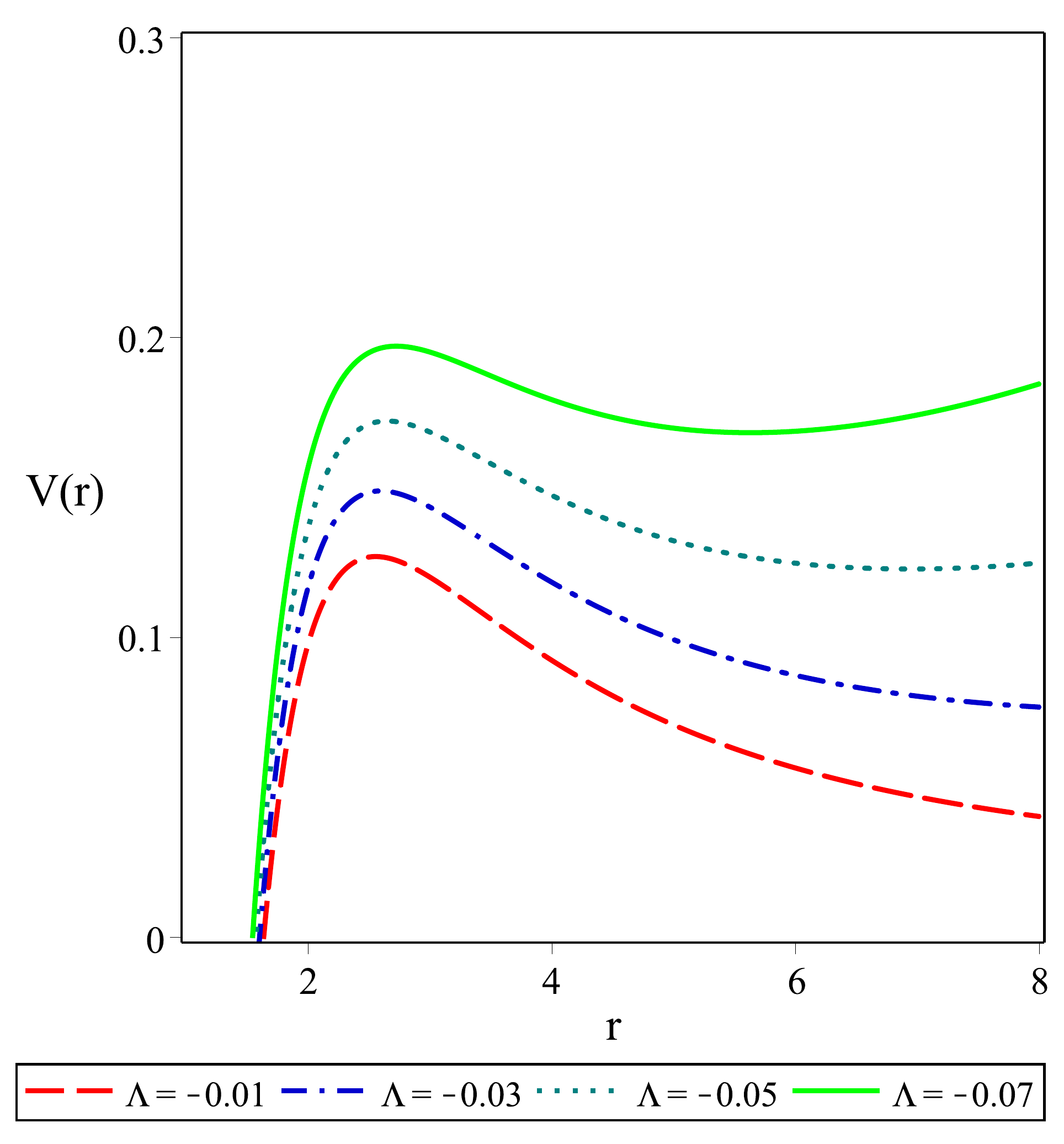}}
       \hspace{1.25cm}
\subfigure[]{
     \label{FigNVd}   \includegraphics[width=0.53\textwidth]{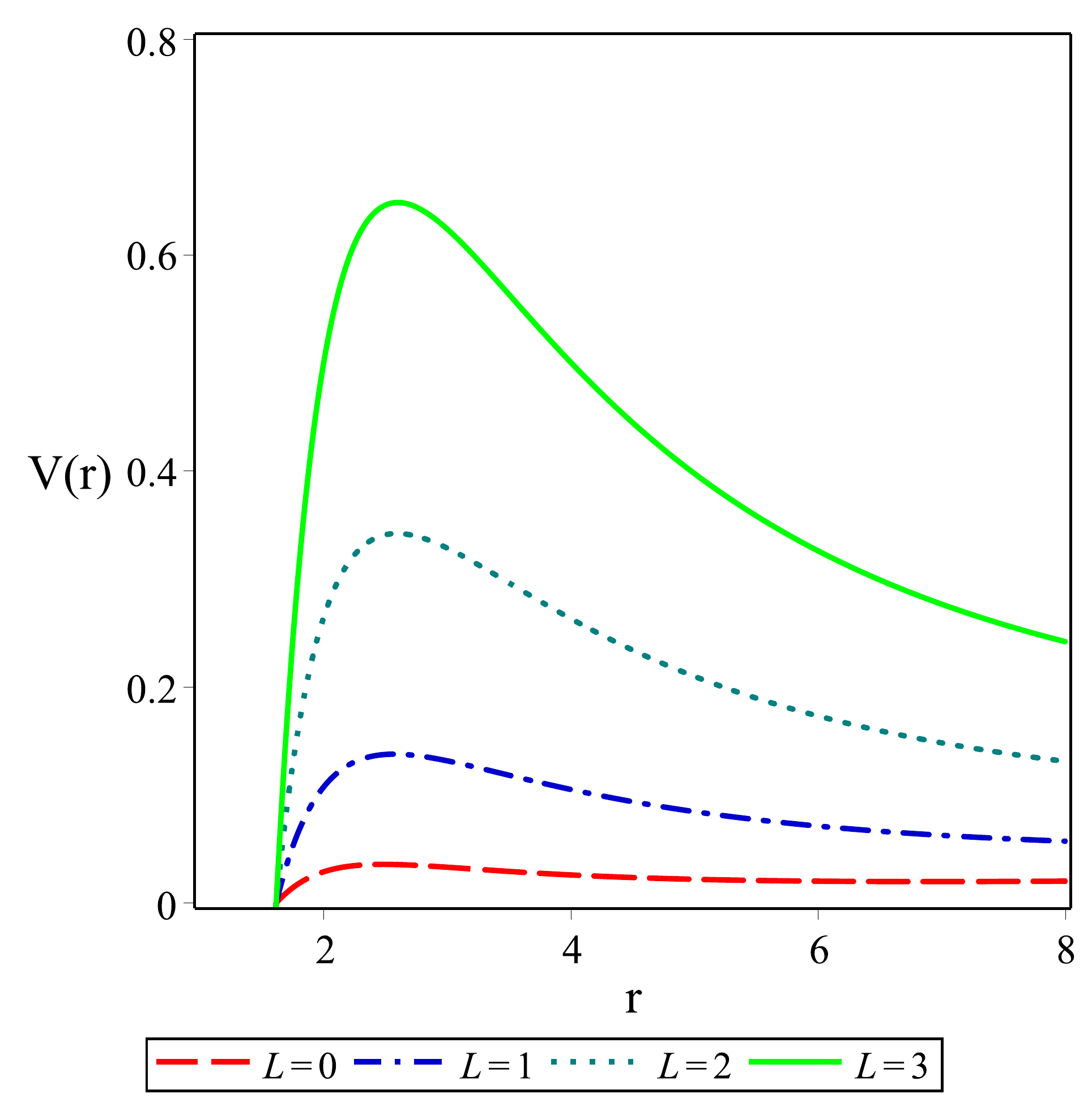}}
     \end{adjustwidth}
\vspace{-6pt}  
\caption{The effective potential for the specific case of $ M=1 $ and different values of $ k $ (Figure~\ref{FigNV}a), $ \alpha $ (Figure~\ref{FigNV}b),  $ \Lambda$ (Figure~\ref{FigNV}c) and  $ L $ (Figure~\ref{FigNV}d). See the text for more~details. (\textbf{a})~$\alpha=0.2$, $ l=1 $ and $\Lambda =-0.02$; (\textbf{b})~$k=0.2$, $ l=1 $ and $\Lambda =-0.02$; (\textbf{c})~$k=\alpha=0.2$ and  $ l=1 $; (\textbf{d}) ~$k=\alpha=0.2$ and $\Lambda =-0.02$.}
\label{FigNV}
\end{figure}

An interesting dynamic quantity is the quasinormal mode, where we consider scalar perturbations by a massless field around the black hole. A~massless canonical scalar field $\Phi$ is described by the Lagrangian density $\mathcal{L}=\frac{1}{2}\left( \partial \Phi \right)^{2}$.
The equation of motion for a scalar field is given by $\nabla^{\mu}\nabla_{\mu} \Phi=0$,
which takes the following form
\begin{equation}
\frac{1}{\sqrt{-g}}\partial_{\mu}(\sqrt{-g}g^{\mu\nu}\partial_{\nu}\Phi)=0.
\end{equation}
Using the method of separation of variables, the~scalar field can be written as
\begin{equation}
\Phi (t, r, \theta, \phi)=\frac{u_{L}(r)}{r}e^{-i\omega t}Y_{Lm}(\theta, \phi),
\end{equation}
where $ Y_{Lm}(\theta, \phi) $ are the standard spherical harmonics. The~radial part of the wave equation satisfies an ordinary second order linear differential equation as follows
\begin{equation}
\frac{d^{2}u_L}{dx^{2}}+ \left[ \omega^{2} - V(x) \right]u_L=0,
\end{equation}
where $ x=\int \frac{dr}{f(r)} $  represents the tortoise coordinates. The~corresponding effective potential barrier is given by
\begin{equation}
V(r)=f(r)\left[ \frac{f^{\prime}(r)}{r}+\frac{L(L+1)}{r^{2}}\right]
\label{EqEV} ,
\end{equation}
where the prime denotes differentiation with respect to $  r$, while $L \geq 0$ is the angular degree, and~$f(r)$ is given by Equation~(\ref{Eqsol1}).  For~scalar and electromagnetic fields, the~effective potentials have the form of a positive definite potential barrier with a single maximum. In~Figure~\ref{FigNV}, the~effective potential is plotted to display how it changes with the NED parameter, the~coupling constant, the~cosmological constant and the spherical harmonic index $ L $. As~we see the height of the potential increases when $ k $, $ \alpha $ and $ L $  increase. The~cosmological constant $\Lambda $ has a decreasing effect on the maximum height of the potential (see Figure~\ref{FigNV}c). A~significant point about this parameter is that the condition of the single maximum is satisfied just for a large cosmological constant ($ \Lambda\geq-0.02 $).

\vspace{-9pt}

The quasinormal modes were studied with different methods over the last decades, as~mentioned in the Introduction. Here, we employ the semi-analytical WKB approximation which was first used in the 1980’s~\cite{SIyer,QNM2} and then extended to 6th order~\cite{QNM3}, and~to 13th order~\cite{Konoplya1}. It is worthwhile to mention that increasing the WKB order does not always lead to a better approximation for the frequency, so we consider the 3rd order expansion for the purpose of our work, which leads to
\begin{eqnarray}
\omega &=& \left\lbrace V+\frac{V_{4}}{8V_{2}}\left(\nu^{2}+\frac{1}{4} \right) - \left( \frac{7+60\nu^{2}}{288} \right)\frac{V_{3}^{2}}{V_{2}^{2}} 
 + i \nu \sqrt{-2V_{2}}\left[ \frac{X_{1}}{2V_{2}} -1\right] \right\rbrace _{r=r_{0}}^{\frac{1}{2}}, \label{EqNV1}
\end{eqnarray} 
where
\begin{eqnarray}
X_{1} &=& \frac{5V_{3}^{4}(77+188\nu^{2})}{6912V_{2}^{4}}
-\frac{V_{4}V_{3}^{2}(51+100\nu^{2})}{384V_{2}^{3}} 
+\frac{V_{4}^{2}(67+68\nu^{2})}{2304V_{2}^{2}}
	\nonumber \\
&&+\frac{V_{5}V_{3}(19+28\nu^{2})}{288V_{2}^{2}}
+\frac{V_{6}(5+4\nu^{2})}{288V_{2}},
\end{eqnarray} 
with $ \nu =n+\frac{1}{2} $ and $ n $ is the overtone number. $ V_{j} $ and $ r_{0} $ represent, respectively,  the~$j$-th derivative of the potential $ V $ and the place in which the height of the potential is maximum. The~interesting fact regarding $r_{0}$ is that it exactly matches the photon sphere radius $r_{p}$ \cite{QNM5}. Expanding the relation (\ref{EqNV1}), we obtain at the eikonal regime $\omega = \omega_{R} -i\omega_{I}$, where
\begin{equation}
\omega_{R}= L\frac{\sqrt{f(r)}}{r}\bigg\vert_{r=r_{0}}+\frac{\sqrt{f(r)}}{2r}\bigg\vert_{r=r_{0}}+\mathcal {O}(L^{-1})\,,
\label{EqQF1a}
\end{equation}
and
\begin{eqnarray}
\omega_{I} &=& \frac{\nu}{\sqrt{2}}\frac{\sqrt{f(r)}}{r}\bigg\vert_{r=r_{0}}\sqrt{6rf^{\prime}-6f-r^{2}f^{\prime\prime}-r^{2}f^{-1}f^{\prime ^{2}}}\Big\vert_{r=r_{0}}
	\nonumber \\
&& \qquad \qquad +\mathcal {O}(L^{-1}),
\label{EqQF1b}
\end{eqnarray}
respectively.

According to the condition $dV/dr\vert_{r=r_{0}} =0 $, one finds that
\begin{equation}
\frac{d}{dr}\left( \frac{f(r)}{r^{2}}\right) \bigg\vert_{r=r_{0}} =0.
\label{EqCV}
\end{equation}
Using Equation~(\ref{EqCV}), the~relations (\ref{EqQF1a}) and (\ref{EqQF1b}) reduce to
\begin{eqnarray}
\omega_{R}&=& r_{s}^{-1}\left( L+\frac{1}{2}+\mathcal {O}(L^{-1})\right) ,
 \nonumber\\
\omega_{I}&=& \frac{2\nu +1}{2\sqrt{2}}r_{s}^{-1}\sqrt{2f-r^{2}f^{\prime\prime}}+\mathcal {O}(L^{-1}),
\label{EqOMI}
\end{eqnarray}
respectively.

Taking Equation~(\ref{EqOMI}) into account, we are in a position to probe how the black hole parameters affect quasinormal frequencies. It is evident that the real part of the modes is proportional to the angular degree, while the imaginary part depends on  the overtone number only. {As already mentioned, the~sign of the imaginary part indicates if the mode is stable or unstable. Our findings show that the imaginary parts of the QNMs are always negative, revealing that the system is stable under perturbations. In~Ref.~\cite{EGB12a}, the~authors considered 4D charged EGB black holes (which is the limit case of the solution (\ref{Eqsol1}) for the case  $r\gg k$ in the absence of the cosmological constant) and investigated the stability/instability of the system through superradiant effects in detail. To~investigate QNMs, one can consider ingoing waves near the event horizon and outgoing waves at infinity. If~the reflected wave amplitude is larger than the incident amplitude, the~wave is amplified. This phenomenon is called superradiance. To~trigger the instability, there should be an effective potential well outside the event horizon to trap the reflected wave from the near-horizon region. In~Ref.~\cite{EGB12a}, it was shown that superradiance is the necessary but not sufficient condition for instability. In~fact, to~confirm the stability of the system, one must ensure that the imaginary part of the frequency is negative. By~comparison, we found that our results are similar to the results obtained in Ref.~\cite{EGB12a}.}

In Figure~\ref{FigNQ},  we show the real and the imaginary part of the frequencies
vs the GB coupling constant for three
different values of the NED parameter, namely, \mbox{$k = 0.1, 0.2,  0.3$} (\mbox{Figure~\ref{FigNQ}a,b}) and three
different values of the cosmological constant $\Lambda = -0.01, -0.02,  -0.03$ (Figure~\ref{FigNQ}c,d). We see that as the GB parameter increases the real part of the QNM frequency grows monotonically, while  the absolute value of the imaginary part of $ \omega $ reduces. This shows that when the coupling constant gets stronger, the~scalar perturbations oscillate more rapidly and due to the decreasing imaginary part, they decay  more slowly. Figure~\ref{FigNQ}a shows how $ Re (\omega) $ changes under varying the NED parameter. Evidently,  the~curves shift upwards by increasing  $ k $, meaning that the real part of the QNMs is an increasing function of this parameter. Whereas, its effect is opposite on $ \vert Im (\omega)\vert $ (see Figure~\ref{FigNQ}b). In~fact, the~effect of the NED parameter on the QNM frequency is similar to that of GB parameter. \mbox{In~Figure~\ref{FigNQ}c,d}, we explore the impact of the cosmological constant on the quasinormal frequency. We verify that this parameter has a decreasing contribution on both of the real and imaginary parts of QNM. This reveals the fact that the scalar perturbations have less energy for oscillations and decay slower in a higher curvature~background.

\vspace{-6pt}
\begin{figure}[H]
\begin{adjustwidth}{-\extralength}{0cm}
\centering%
\subfigure[]{
\label{FigNQa}        \includegraphics[width=0.55\textwidth]{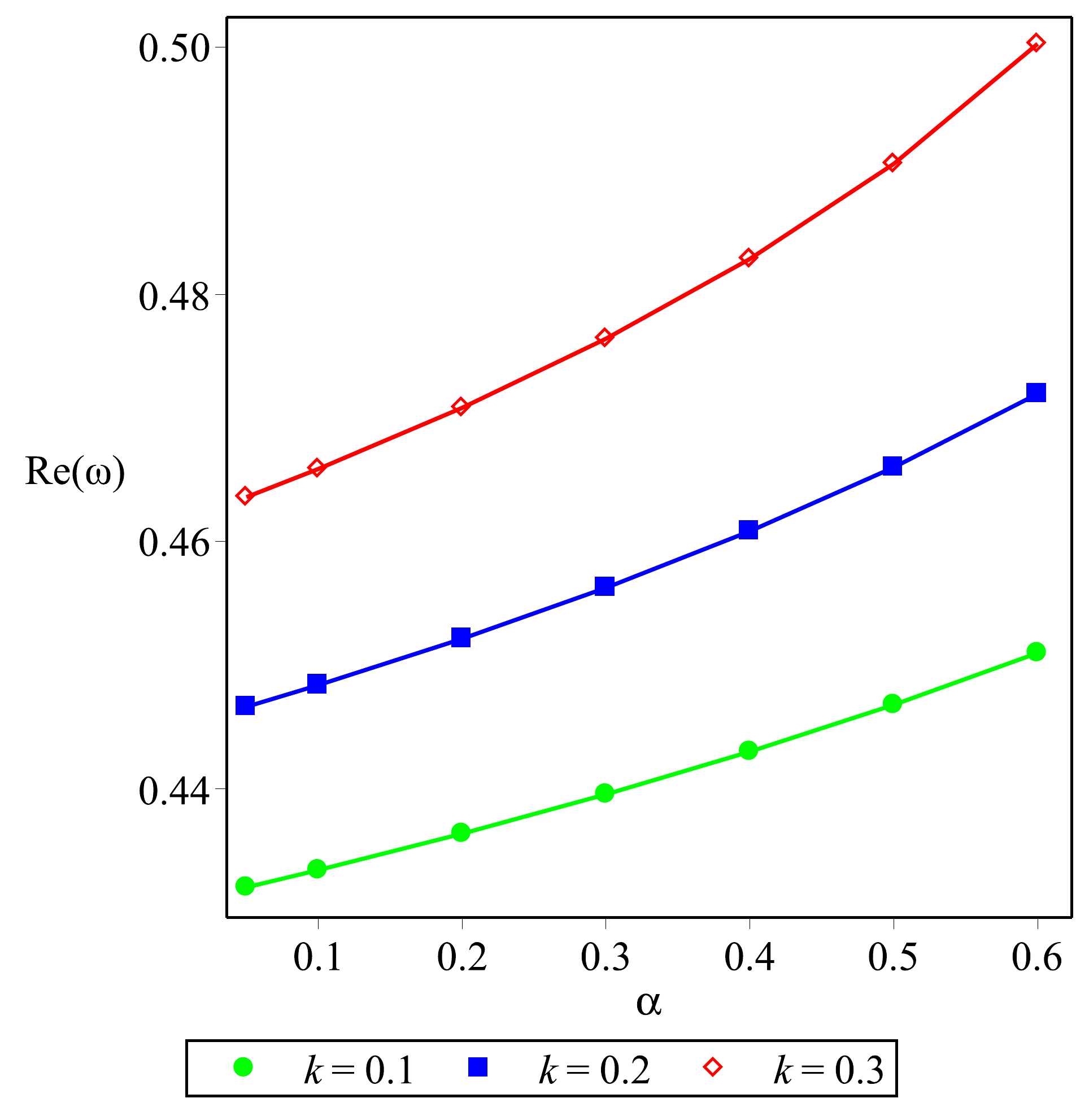}}
     \hspace{1.25cm}
\subfigure[]{
\label{FigNQb}        \includegraphics[width=0.56\textwidth]{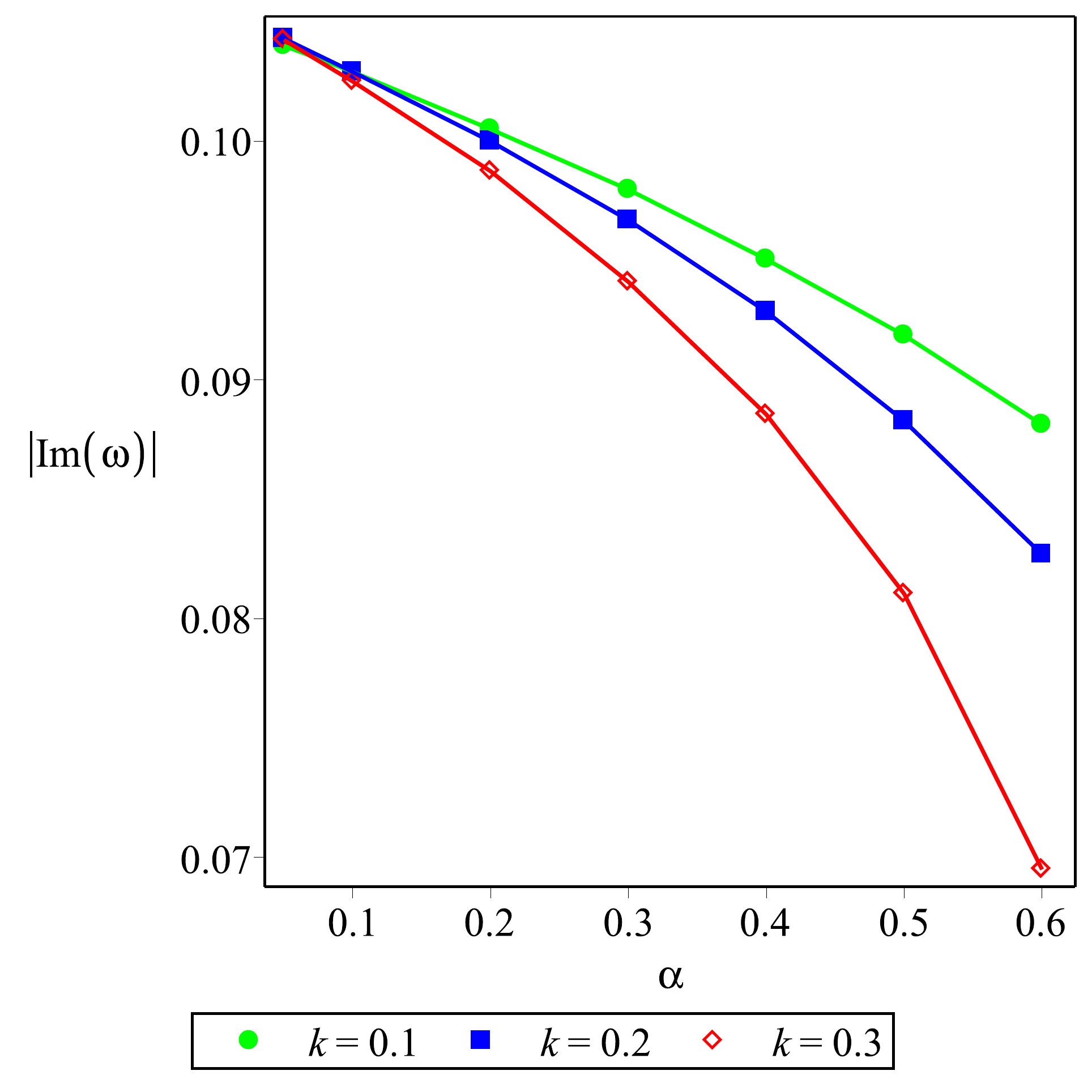}}
\end{adjustwidth}
\caption{\textit{Cont}.}
\label{FigNQ}
\end{figure}

\begin{figure}[H]\ContinuedFloat
\setcounter{subfigure}{2}
\begin{adjustwidth}{-\extralength}{0cm}
\centering%
\subfigure[]{
\label{FigNQc}        \includegraphics[width=0.55\textwidth]{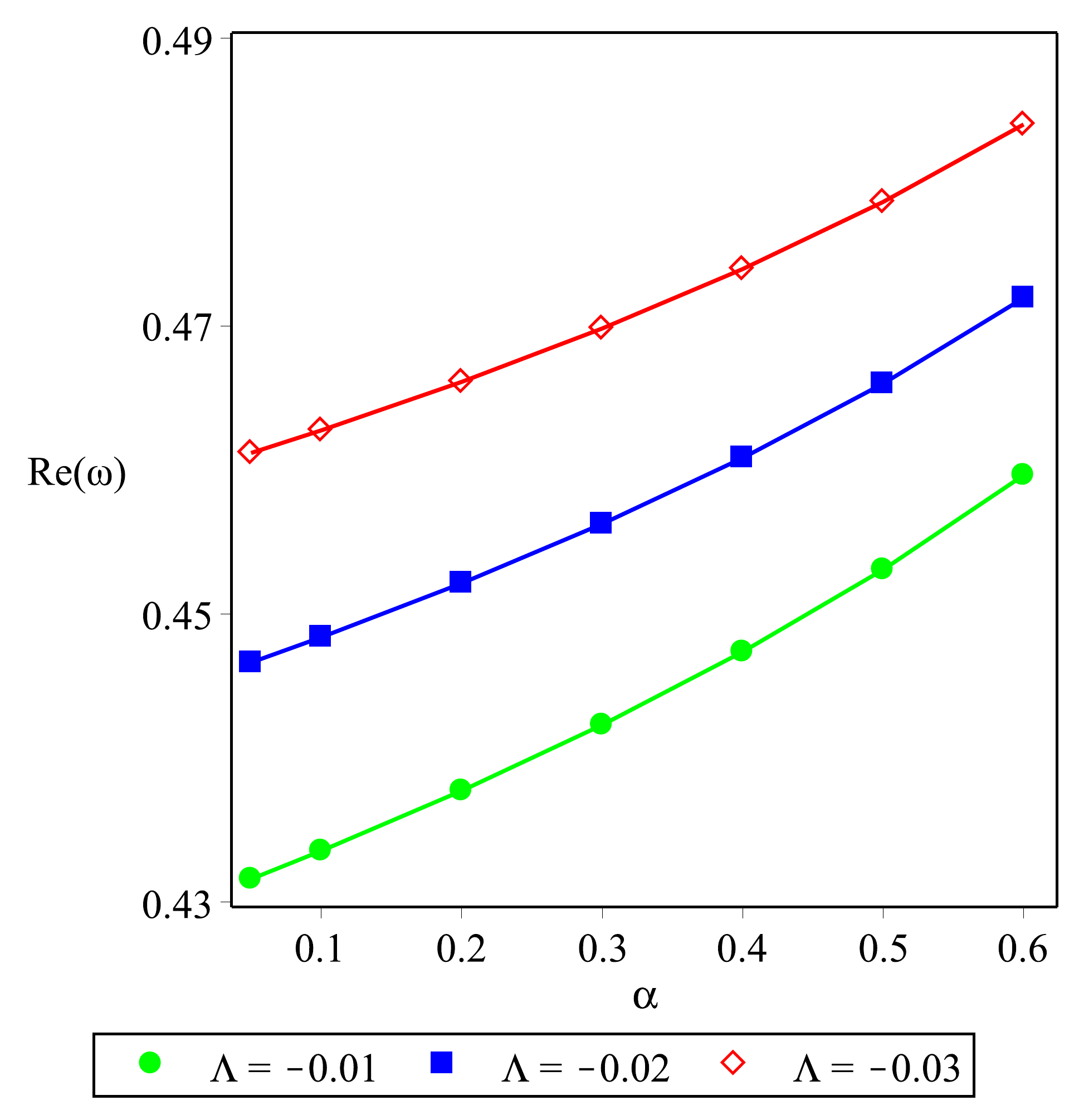}}
     \hspace{1.25cm}
\subfigure[]{
\label{FigNQd}        \includegraphics[width=0.56\textwidth]{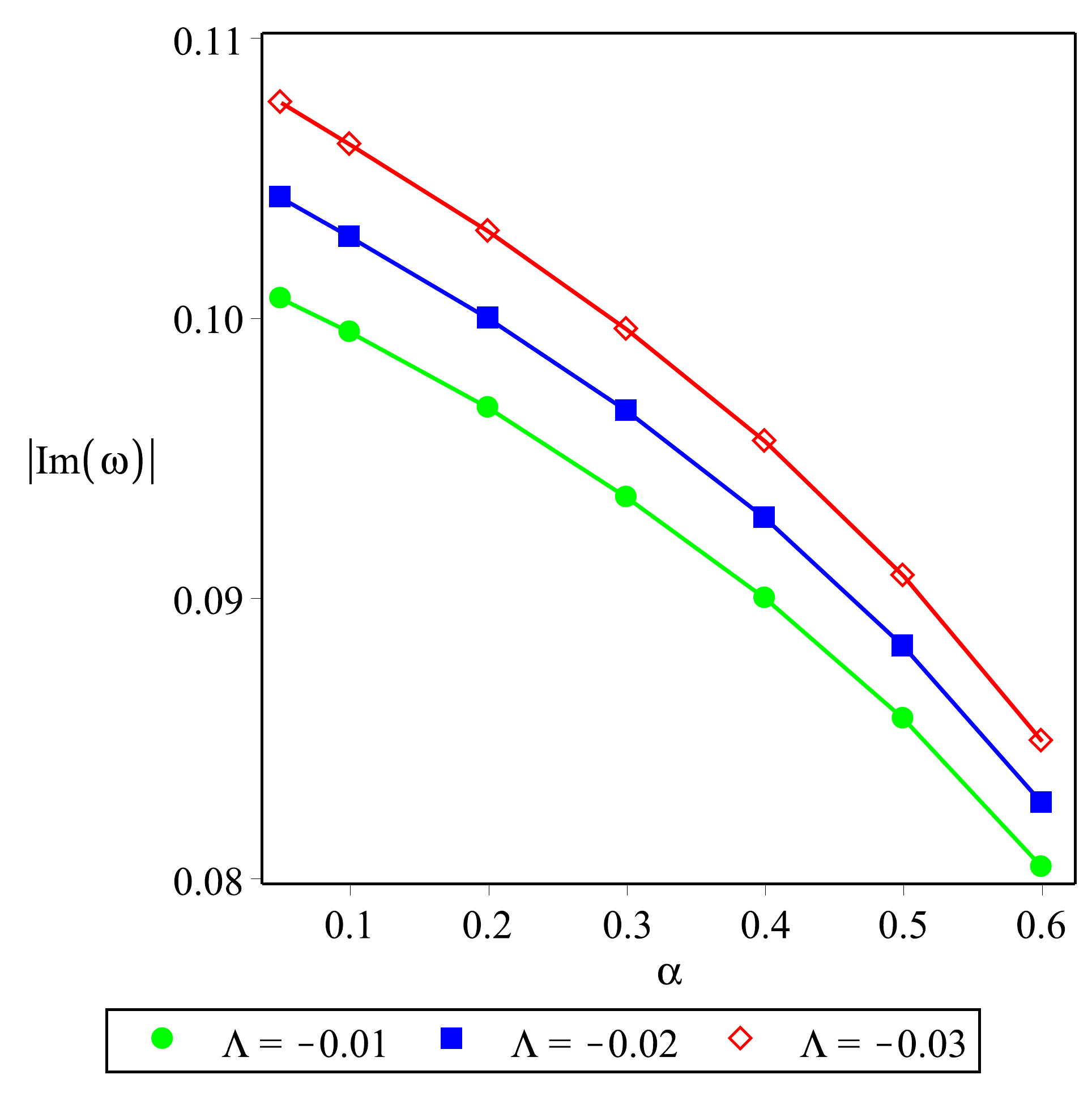}}
\end{adjustwidth}
\vspace{-6pt}     
\caption{The behavior of $Re (\omega)$ and $Im (\omega)$ with the GB parameter for $ M=1 $, $ n=0 $, $ L=2 $ and different values of $ k $ (Figure~\ref{FigNQ}a,b) and  $ \Lambda$ (Figure~\ref{FigNQ}c,d). See the text for more~details. (\textbf{a})~$\Lambda =-0.02$; (\textbf{b})~$\Lambda =-0.02$; (\textbf{c})~$k=0.2$; (\textbf{d})~$k=0.2$.}
\label{FigNQ2}
\end{figure}

\section{Summary and Concluding~Remarks}\label{sec:conclusion}

Among the higher curvature gravitational theories, the~recently proposed novel 4D EGB gravity has received an extensive and intensive interest, especially in terms of applications in regularized 4D EGB black hole solutions. Motivated by the importance of such black holes, in~this work, we presented a study in the context of dynamic optical features of AdS black holes coupled to nonlinear electrodynamics,  including the shadow size, the~energy emission rate, the~deflection angle and quasinormal modes. We first investigated  the photon sphere and the shadow observed by a distant observer and discussed how the black hole parameters affect  them. We found that both the photon sphere radius and the shadow size decrease with an increasing GB parameter. Studying the impact of the cosmological constant, we verified that the shadow size shrinks by increasing this parameter which is opposite to the behavior observed in the photon sphere radius. Regarding the NED parameter, it had a decreasing contribution on both the radius of the photon sphere and shadow similar to the GB coupling~constant.

Then, we continued by studying the energy emission rate and explored the effect of the black hole parameters on the radiation process. The~results showed that as the coupling constants and the cosmological constant increase, the~emission of particles around the black hole decreases. This revealed the fact that the radiation rate grows when the effects of the coupling constants get weaker or the background curvature becomes lower. In~other words, the~lifetime of a black hole would be shorter under such conditions.
Furthermore,  we studied the gravitational lensing of light around such black holes.  We found that there is an inverse relationship  between the deflection angle and impact parameter. Thus, the~deflection angle decreases with the increasing of the impact parameter. Both the GB parameter and the cosmological constant exhibited decreasing effects on $ \Theta $, unlike the NED parameter which had an increasing contribution. We also noticed that the variation of the cosmological constant does not have a remarkable effect on the deflection angle. This revealed the fact that varying the background curvature does not affect the path of light significantly. In~contrary to $\Lambda$, the~NED parameter had a significant contribution on $ \Theta $ compared to the other two~parameters.

Finally, we presented a study of the quasinormal modes of scalar perturbation. We found that increasing the coupling constants lead to increasing (decreasing) of the real part (absolute value of the imaginary part) of the quasinormal frequencies. This indicated that as the effect of the coupling constants gets stronger, the~QNMs oscillate faster and decay slower. The~cosmological constant had a decreasing effect on both the real and imaginary parts of the QNMs. This indicates that although the scalar perturbations have less energy for oscillation in a higher curvature background,  they decay slower in such a situation.
Here, we studied the optical features of regular AdS black hole solutions of 4D EGB gravity coupled to exponential nonlinear electrodynamics. There are other nonlinearly charged black solutions in this context as well, such as Born-Infeld charged solution presented in~\cite{EGB5e}. It is also interesting to study such optical properties for these solutions. These issues are now under investigation, and~the results will appear~elsewhere.

As we know, modern observational results have shown that our  universe is expanding with acceleration, indicating the fact that  a static or stationary observer may see a time-dependent shadow. One of the well-known models of dark energy to explain the accelerating expansion is the cosmological constant, which has been the subject of extensive studies in the context of gravitational lensing in recent years~\cite{Sh2,Sh3,Sh4}. It should be noted that although the expansion of the universe was based on a positive cosmological constant, there are also some evidence that show that it may be associated with a negative cosmological constant. For~instance, the~studies related to the Hubble constant data $H(z)$ at low redshift showed that the dark energy density has a negative minimum for certain redshift ranges which can be simply modeled through a negative cosmological constant~\cite{Dutta12}. The~other reason is related to the concept of stability of  the accelerating universe. The~analysis of the de Sitter expanding spacetime with a constant internal space showed that the de Sitter solution would be stable just in the presence of  the negative cosmological constant~\cite{Maedaa}. The~other interesting reason is through supernova data. Although~astrophysical observations from high-redshift supernova have shown that the expansion of the Universe is accelerating due to a positive cosmological constant, the~supernova data themselves derive a negative mass density in the Universe~\cite{Riess,Perlmutter}. According to the results of Ref.~\cite{Farnes}, such a negative mass density is equivalent to a negative cosmological constant. In~fact, the~introduction of negative masses can lead to an anti de Sitter~space. 

It is also remarkable to mention that the phase structure of black holes can be reflected in the shadow size~\cite{M.Zhang} and the corresponding QNMs~\cite{JJing1,XHeB,Shen:2007xk}. So, the~study of these phenomena can provide important information about black hole thermodynamics and is significant from a holographic point of view where the black hole is dual to a system on the AdS boundary. Such a study was behind the aim of the present paper and the issue will be investigated elsewhere. Moreover, from~a cosmological viewpoint, there is a possibility to have an accelerating expansion in the context of theories accompanied by a negative cosmological constant~\cite{Dutta12,Maedaa,Riess,Perlmutter,Farnes,Hartle:2012qb} and thus the present results may also be relevant in such a~case.

\vspace{6pt}


\authorcontributions{All the authors have substantially contributed to the present~work. All authors have read and agreed to the published version of the manuscript.}

\funding{This research was funded by Funda\c{c}\~{a}o para a Ci\^{e}ncia e a Tecnologia (FCT) through the grants Scientific Employment Stimulus contract with reference CEECIND/04057/2017, and UIDB/FIS/04434/2020, No. PTDC/FIS-OUT/29048/2017}

\institutionalreview{Not applicable.}

\informedconsent{Not applicable.}



\acknowledgments{\textls[-10]{M.K.Z. thanks Shahid Chamran University of Ahvaz, Iran for supporting this work under research grant No. SCU.SP1400.37271. F.S.N.L. acknowledges support from the Funda\c{c}\~{a}o para a Ci\^{e}ncia e a Tecnologia (FCT) Scientific Employment Stimulus contract with reference CEECIND/04057/2017. F.S.N.L. also thanks funding from the FCT research grants Nos. UIDP/FIS/\linebreak04434/2020, UIDB/FIS/04434/2020, No. PTDC/FIS-OUT/29048/2017 and No. CERN/FIS-PAR/0037/\linebreak2019.}}

\conflictsofinterest{The authors declare no conflict of~interest.}

\begin{adjustwidth}{-\extralength}{0cm}
\printendnotes[custom] 

\reftitle{References}

\end{adjustwidth}



%


%

\end{document}